\documentclass[]{pasj01}
\usepackage{color}
\usepackage{threeparttable}
\usepackage[figuresright]{rotating}
\usepackage{lscape}
\newcommand{\Rcico}{$R_{\rm [\atom{C}{}{}\emissiontype{I}]/\atom{CO}{}{}}$ }
\newcommand{\ci}{[\atom{C}{}{}\emissiontype{I}](1--0)}
\newcommand{\C}{[\atom{C}{}{}\emissiontype{I}]}

\newcommand{\um}{$\mu$m}
\newcommand{\co}{\atom{CO}{}{12}}
\newcommand{\CO}{\atom{CO}{}{13}}
\newcommand{\hi}{\atom{H}{}{}\emissiontype{I}}

\begin{document} 
\Received{2021/01/15}
\Accepted{2021/03/04}

\title{ Atomic Carbon [\atom{C}{}{}\emissiontype{I}]$(^3P_1-^3P_0)$ Mapping of the Nearby Galaxy M83}

\author{Yusuke \textsc{Miyamoto}\altaffilmark{1,*}%
}
\altaffiltext{1}{National Astronomical Observatory of Japan, 2-21-1 Osawa, Mitaka, Tokyo 181-8588, Japan}
\email{miyamoto.yusuke@nao.ac.jp, mymt.yusuke@gmail.com}

\author{Atsushi \textsc{Yasuda}\altaffilmark{2}}
\altaffiltext{2}{Department of Physics, Graduate School of Pure and Applied Sciences, University of Tsukuba, 1-1-1 Tennodai, Tsukuba, Ibaraki 305-8571, Japan}

\author{Yoshimasa  \textsc{Watanabe}\altaffilmark{3}}
\altaffiltext{3}{Materials Science and Engineering, College of Engineering, Shibaura Institute of Technology, 3-7-5 Toyosu, Koto-ku, Tokyo 135-8548, Japan}

\author{Masumichi  \textsc{Seta}\altaffilmark{4}}
\altaffiltext{4}{School of Science and Technology, Kwansei Gakuin University, 2-1 Gakuen, Sanda, Hyogo 669-1337, Japan}

\author{Nario  \textsc{Kuno}\altaffilmark{2, 5}}
\altaffiltext{5}{Tomonaga Center for the History of the Universe, University of Tsukuba, 1-1-1 Tennodai, Tsukuba, Ibaraki 305-8571, Japan}

\author{Dragan  \textsc{Salak}\altaffilmark{5}}

\author{Shun  \textsc{Ishii}\altaffilmark{1}}

\author{Makoto  \textsc{Nagai}\altaffilmark{1}}

\author{Naomasa  \textsc{Nakai}\altaffilmark{4}}



\maketitle
\begin{abstract}
Atomic carbon (\atom{C}{}{}\emissiontype{I}) has been proposed to be a global tracer of the molecular gas as a substitute for \atom{CO}{}{}, however, its utility remains unproven.
To evaluate the suitability of \atom{C}{}{}\emissiontype{I} as the tracer, 
we  performed \C$(^3P_1-^3P_0)$ (hereinafter \ci) mapping observations of the northern part of the nearby spiral galaxy M83 with the ASTE telescope and
compared the distributions of \ci~with \atom{CO}{}{} lines (\atom{CO}{}{}(1--0), \atom{CO}{}{}(3--2), and \atom{CO}{}{13}(1--0)), \atom{H}{}{}\emissiontype{I}, and infrared (IR) emission (70, 160, and 250$~\mu$m).
The \ci~distribution in the central region is similar to that of the \atom{CO}{}{} lines,  
whereas \ci~in the arm region is distributed outside the CO.
We examined the  dust temperature, $T_{\rm dust}$, and dust mass surface density, $\Sigma_{\rm dust}$, 
by fitting the IR continuum-spectrum distribution with a single temperature modified blackbody.
The distribution of $\Sigma_{\rm dust}$ shows a much better  consistency with the integrated intensity of \atom{CO}{}{}(1--0) than with that of \ci, 
indicating that \atom{CO}{}{}(1--0) is a good tracer of the cold molecular gas.
The spatial distribution of the \C~excitation temperature, $T_{\rm ex}$, was examined using the intensity ratio of the two \C~transitions.
An appropriate $T_{\rm ex}$ at 
the central, bar, arm, and inter-arm regions 
yields a constant [\atom{C}{}{}]/[\atom{H}{}{}$_2$] abundance ratio of  $\sim7 \times 10^{-5}$ within a range of 0.1~dex in all regions. 
We successfully detected weak \ci~emission, 
even in the inter-arm region, in addition to the central, arm, and bar regions,  using spectral stacking analysis. 
The stacked intensity of \ci~is found to be strongly correlated with $T_{\rm dust}$.
Our results indicate that the atomic carbon is a photodissociation product of \atom{CO}{}{}, 
and consequently, 
compared to \atom{CO}{}{}(1--0), 
\ci~is less reliable in tracing the bulk of "cold" molecular gas in the galactic disk.

\end{abstract}



\section{Introduction}
Molecular hydrogen (\atom{H}{}{}$_2$) is a major constituent of the interstellar medium in galaxies. 
However, direct detection of cold H$_2$ is difficult because \atom{H}{}{}$_2$ lacks a permanent dipole moment due to its symmetric structure.
Alternatively, the amount of \atom{H}{}{}$_2$ has been indirectly estimated by observing \atom{CO}{}{}, 
which is the second most abundant species in molecular clouds.
\atom{CO}{}{} has a weak permanent dipole moment and a ground rotational transition with a low excitation energy of $\sim 5.5$~K. 
Owing to this low energy and critical density ($n_{\rm cr}\sim10^3$~cm$^{-3}$, however, $n_{\rm cr}$ is further reduced by radiative trapping due to its large optical depth), \atom{CO}{}{} is easily excited even in cold molecular clouds. 
Meanwhile, the large optical depth makes it difficult to estimate the molecular gas mass.
Furthermore, in a low-metal environment, the \atom{CO}{}{} lines are not necessarily suitable for tracing the amount of molecular gas because of \atom{CO}{}{} depletion that is caused by the lack of heavy elements and by the photo-dissociation due to the interstellar radiation field as a result of poor (self)shielding (e.g., \cite{Israel1997}, \cite{Leroy2007}).

The forbidden fine-structure transition lines of atomic carbon, namely, [\atom{C}{}{}\emissiontype{I}]$(^3P_1-^3P_0)$ and [\atom{C}{}{}\emissiontype{I}]$(^3P_2-^3P_1)$, hereinafter \ci~and \C(2--1), respectively, are optically thin in most cases (e.g., \cite{Ikeda2002}); 
further, they have excitation energies of 23.6~K and 62.5~K, respectively, and  the critical density of  \ci~($n_{\rm cr} \approx 10^3$~cm$^{-3}$) is similar to that of \atom{CO}{}{}(1--0).
In addition, 
although the classical photodissociation region (PDR) models (\cite{Tielens1985}, \cite{Hollenbach1991}) expect 
the atomic carbon to exist predominantly in the thin layer near the surface of a homogeneous static molecular cloud that is exposed to UV radiation, 
\ci~mapping observations of galactic molecular clouds have shown that  the distribution of \ci~is similar to that of  low-$J$ \atom{CO}{}{}  (and \atom{CO}{}{13}) lines (e.g.,\cite{Tauber1995}, \cite{Ikeda2002}, \cite{Shimajiri2013}).
The co-existence of \ci~with \atom{CO}{}{} can be explained by introducing density inhomogeneity in the densities of molecular clouds in the classical PDR, because it allows the external radiation to penetrate deeply into the cloud  (e.g., \cite{Spaans1996}). 
Recent refined simulations such as those considering the turbulence of the clouds \citep{Glover2015} or the influence of cosmic rays (e.g., \cite{Bisbas2015}, \cite{Papadopoulos2018})  have confirmed the widespread distribution of \ci~in the clouds. 
The results of both these observations and theories have 
encouraged the understanding 
that \C~could be a molecular gas tracer alternative to low-$J$ \atom{CO}{}{}.

Recently, owing to the availability of the Atacama Large Millimeter/submillimeter Array (ALMA) and Hershel/SPIRE, 
the relation between [\atom{C}{}{}\emissiontype{I}] and \atom{CO}{}{} for extra-galaxies has been actively surveyed 
to confirm  the utility of [\atom{C}{}{}\emissiontype{I}] as a molecular gas tracer 
[e.g., 
 \citet{Israel2015}, \citet{Jiao2017}, and \citet{Jiao2019} for local and (ultra)luminous infrared galaxies ((U)LIRGs); 
\citet{Valentino2018} for high-redshift galaxies].
\citet{Jiao2017} and \citet{Jiao2019} have demonstrated a strong (nearly linear) relation between [\atom{C}{}{}\emissiontype{I}] and \atom{CO}{}{}(1--0) line luminosities for (U)LIRGs and local galaxies, and concluded that \ci~is likely a good molecular gas mass tracer.
In contrast,
\citet{Israel2015}  argued that \ci~may trace dense clouds rather than a diffuse gas, and consequently, it is not a good tracer of the bulk of molecular gas.
Moreover, through  observations of the individual galaxies, a careful treatment for using \ci~as an alternative  molecular gas tracer to \atom{CO}{}{} is cautioned because of the spatial variance of the  [\atom{C}{}{}\emissiontype{I}]--\atom{CO}{}{} relation (e.g., \cite{Salak2019}, \cite{Saito2020}).
Thus, whether \C~can be a \atom{H}{}{}$_2$ tracer as a substitute for \atom{CO}{}{} is  under debate.
Notably,  
the investigations of the relation between \ci~and \atom{CO}{}{} 
have been mainly biased toward \atom{CO}{}{}-bright sources, such as star-forming clouds, the centers of galaxies, and (U)LIRGs,
thus far.
To understand the  \atom{C}{}{}\emissiontype{I}--\atom{CO}{}{} relation and the utility of  \ci~as a molecular gas tracer, 
it is necessary to compare \atom{C}{}{}\emissiontype{I} and \atom{CO}{}{} with 
\atom{H}{}{}$_2$ estimated by an independent method, independent of the brightness of \atom{CO}{}{}.

Infrared (IR) dust emission, which is optically thin over most regions of normal galaxies is another tool to estimate the molecular gas surface density, $\Sigma_{\rm mol}$.
Under the assumptions that dust and gas are well mixed and the gas-to-dust ratio ($\textrm{GDR}$) is fixed in the atomic and molecular phases, 
the dust mass surface density, $\Sigma_{\rm dust}$, measured with the IR emission can be converted to the total gas surface density, $\Sigma_{\rm gas}$, using the $\textrm{GDR}$.
Then, $\Sigma_{\rm mol}$ can be obtained by subtracting the contribution of the atomic gas surface density, $\Sigma_{\rm atom}$, based on the \atom{H}{}{}\emissiontype{I} measurement from $\Sigma_{\rm gas}$.
However, the determination of $\textrm{GDR}$ is an unresolved issue.
Considering the following equation, 
\begin{eqnarray}
\Sigma_{\rm gas} &=& \textrm{GDR}~\Sigma_{\rm dust} \nonumber\\
&=& \Sigma_{\rm atom} + \alpha_{\rm \atom{CO}{}{}}I_{\rm \atom{CO}{}{}},
\label{eq:gas}
\end{eqnarray}
where $\alpha_{\rm \atom{CO}{}{}}$ is the \atom{CO}{}{}-to-\atom{H}{}{}$_2$ conversion factor and $I_{\rm \atom{CO}{}{}}$ is the measured \atom{CO}{}{} integrated intensity, 
\citet{Leroy2011} proposed a technique to simultaneously measure the $\textrm{GDR}$ and $\alpha_{\rm \atom{CO}{}{}}$ under the assumption that the $\textrm{GDR}$ is constant over a region of the galaxy.
Applying this technique to resolved galaxies allows us to compare \C~and \atom{CO}{}{} with the dust-based $\Sigma_{\rm mol}$ for each galactic structure.

M83 is an ideal target to investigate the \C--\atom{CO}{}{} relation for the galactic structures,  
because it is one of the nearest (4.5 Mpc, \cite{Thim2003}) spiral galaxies that is nearly face-on (inclination=\timeform{24D}; \cite{Comte1981}) and hosts prominent galactic structures (a bar and spiral arms). 
Moreover, multi-wavelength images  for \atom{H}{}{}\emissiontype{I}, \atom{CO}{}{}, IR, and optical observations can be acquired from the archives.
The basic parameters of M83 adopted in this paper are summarized in Table~\ref{tab:para}.

\begin{threeparttable}
  \caption{Parameters of M83}%
  \label{tab:para}
  \begin{tabular}{ll}
      \hline
      Parameter & Value \\ 
      \hline
      $\alpha_{\rm J2000.0}$\tnote{a}  		& \timeform{13h37m00s.48}\\
      $\delta_{\rm J2000.0}$\tnote{a}  		& \timeform{-29D51'56''.48}  \\
      Distance\tnote{b} 				& 4.5~Mpc  \\
      Position angle\tnote{c} 			& \timeform{225D}  \\
      Inclination angle\tnote{c} 			& \timeform{24D}  \\
      $R_{25}$\tnote{d} 				&6\farcm44 \\
      Systemic velocity (LSR)\tnote{e} 		& $514\pm5$~km~s$^{-1}$  \\
      Linear scale (Beam size $=18\farcs9$) & $\sim 412$~pc \\
      \hline      
    \end{tabular}
    \begin{tablenotes}
    \item[a] \cite{Sukumar1987}
    \item[b] \cite{Thim2003} 
    \item[c] \cite{Comte1981}
    \item[d] \cite{de1991}
    \item[e] \cite{Kuno2007}
    \end{tablenotes}
\end{threeparttable}

\section{Observations and Analyses}
\subsection{CI data}
\label{chap:obs}
\begin{figure}[th]
 \begin{center}
  \includegraphics[width=\linewidth]{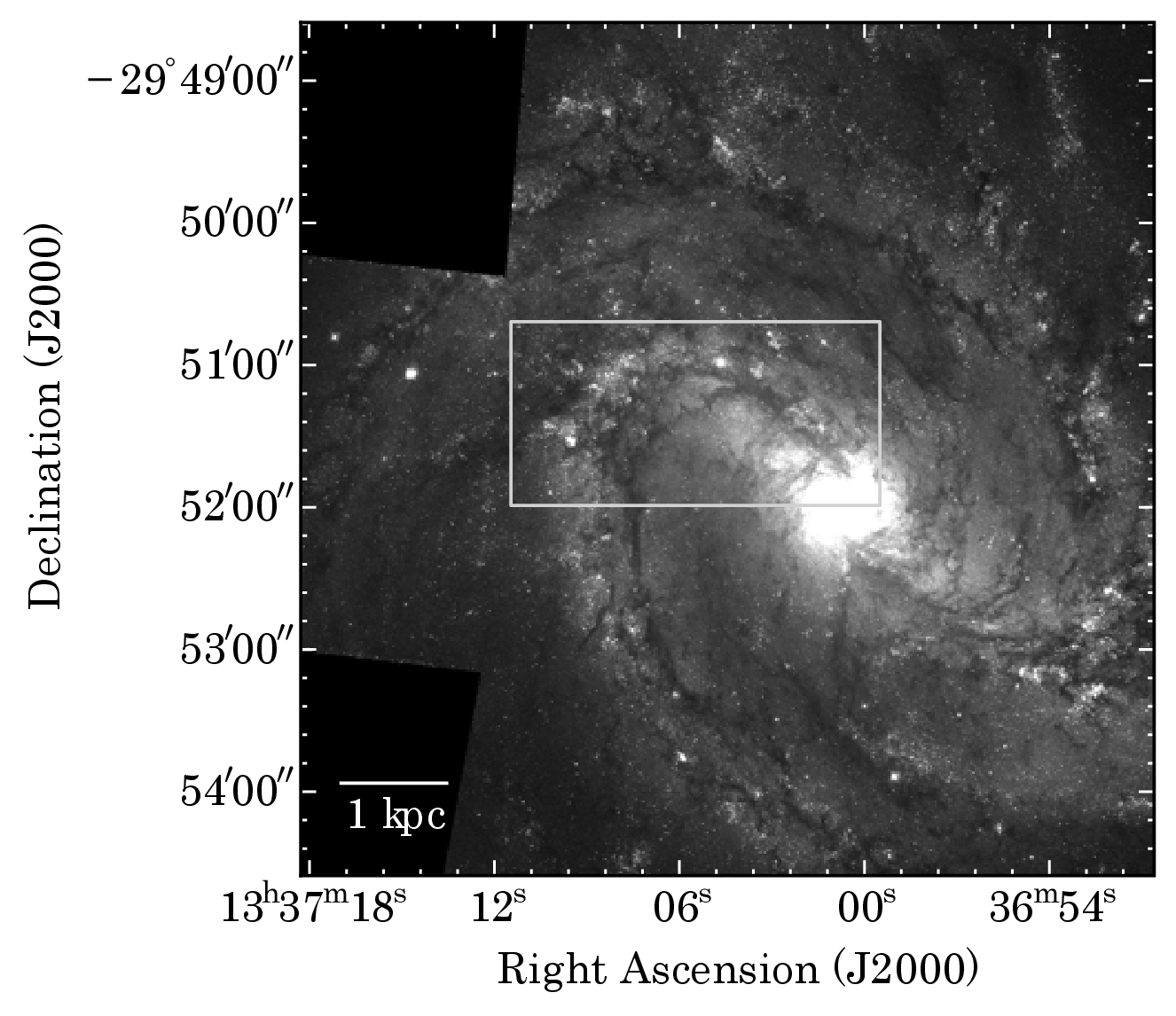}
 \end{center}
 \caption{ 
 Observed $\timeform{150''}\times \timeform{72''}$ area (gray line) using the ASTE superposed on a V-band image of M83 retrieved from the data archive of the HST/WFC3 Early Release Science Program.
 }
 \label{fig:m83}
\end{figure}
Observations of M83 in \ci~($\nu_{\rm rest}=492.160651$~GHz) were carried out from July to August of 2017 and  July 2019 using the ASTE 10-m telescope.
The observations were performed in the on-the-fly (OTF) mapping mode \citep{Sawada2008}.
The observed area covered a $\timeform{150''}\times \timeform{72''}$ rectangular region with PA$=\timeform{0D}$ and its center offset from the galactic center $(\alpha, \delta)_{\rm J2000.0} =$ (\timeform{13h37m00s.48}, \timeform{-29D51'56''.48}) \citep{Sukumar1987}
was ($\Delta \alpha, \Delta \delta$)=($\timeform{65''},\timeform{36''}$) so that  
the  area could include the galactic center, 
the northern galactic bar, 
and a part of the northern spiral arm (Figure~\ref{fig:m83}). 
The effective beam size was 18\farcs{9} at 492~GHz for the OTF observations, corresponding to 412~pc at a distance of 4.5 Mpc \citep{Thim2003}, which allowed us to distinguish the galactic structures, such as the bar and arm.
The scan separation, perpendicular to the scan direction, was \timeform{6"} and the mapping was performed in two orthogonal scan directions so that the noise temperatures in each direction were as even as possible to remove any effects of scanning noise. 
Two points with \timeform{10'} offset from the map center  in the direction of the declination  were used as the off-source positions.
The pointing toward a Mira-type variable star W~Hya 
was checked every 1--2~hr by observing \atom{CO}{}{}(3--2), and the accuracy was  better than \timeform{4"}.

The utilized frontend was a two-sideband dual-polarization heterodyne receiver of  ASTE Band~8 and the backend was an FX-type spectrometer, WHSF, with a  total bandwidth of 2048~MHz. 
The total number of channels was 2048, with a spectral resolution of 1~MHz (0.6~km~s$^{-1}$ at 492~GHz), 
and a velocity coverage of 1248~km~s$^{-1}$.
The  system noise temperature at 492~GHz ranged from 500 to 2500~K.
The line intensity was calibrated by the chopper wheel method, yielding an antenna temperature, $T_{\rm A}^{\ast}$, corrected for both atmospheric and antenna ohmic losses \citep{ulich1976}. 
In this study, we used the main beam brightness temperature $T_{\rm mb}\equiv T_{\rm A}^{\ast}/\eta_{\rm mb}$, with the main beam efficiency of the antenna, $\eta_{\rm mb} =$ 0.45.
The antenna temperature of 1~K corresponds to 127.2 Jy at 492~GHz.
The absolute intensity and the variation in the main beam efficiency was checked by observing M~17 in [\atom{C}{}{}\emissiontype{I}] and 
comparing the standard spectra of  M~17 \citep{white1991}.
The uncertainty of the efficiency was estimated to be better than 20\%.

We used an auto-reduction system, {\tt COMING ART} \citep{sorai2019}, for data reduction, 
based on Nobeyama OTF Software Tools for Analysis and Reduction ({\rm NOSTAR}), developed by Nobeyama Radio Observatory. 
The 
{\tt COMING ART} system 
flagged and removed poor quality data after applying a linear baseline fitting, and then 
performed the basket weaving procedure to reduce the scanning effect \citep{emerson1988}.
Before preparing the final  [\atom{C}{}{}\emissiontype{I}] cube data with \timeform{6"} spacing and a velocity width of 10~km~s$^{-1}$, a cubic polynomial baseline fitting was applied in the emission-free range to optimize the zero level in each spectrum. 
The resultant rms noise level was typically 25~mK in the $T_{\rm mb}$ scale.

\subsection{\atom{CO}{}{} data}
\label{sec:co}
We retrieved the \atom{CO}{}{12}(1--0), \atom{CO}{}{13}(1--0),  and  \atom{CO}{}{12}(3--2) data from the ALMA archive (projects 2012.1.00762.S and 2015.1.01593.S, PI: Hirota), all of which contained data obtained with the 12-m, 7-m, and total power (TP) arrays.
To obtain images matching with the spatial resolution of [\atom{C}{}{}\emissiontype{I}], we used the data obtained with the 7-m and TP arrays for  \atom{CO}{}{12}(1--0) and \atom{CO}{}{13}(1--0) and the TP array for \atom{CO}{}{12}(3--2)\footnote{For comparison, \atom{CO}{}{}(3--2) data obtained with the 12-m and 7-m arrays were  calibrated 
in the same manner  as that described in the text 
and combined with the TP data through the {\tt Feather} algorithm after imaging the concatenated data (Figure~\ref{fig:results}(h)).}.
We processed the data with the observatory-provided calibration scripts through CASA \citep{mcmullin2007}. 
The calibration was carried out in CASA version 4.5.3 or 4.7.2, following the manuals provided by the  observatory. 
After inspecting the calibrated data, we subtracted the continuum emission for each line determined at the emission-free channels.
Imaging of the interferometric data  was performed in {\tt tclean} in CASA version 5.6.1, where visibilities in  baselines longer than $\sim 12$~k$\lambda$ in the uv-plane were tapered to maximize sensitivity to the extended structures.
The calibrated TP data in each line were imaged in {\tt sdimaging}.
Finally, we combined these data with the {\tt Feather} algorithm in CASA.
All of the final data were spatially smoothed to match the resolution of the [\atom{C}{}{}\emissiontype{I}]  data (\timeform{18.9"}).

\subsection{\hi~data}
To trace the atomic hydrogen gas surface density of M83,
we retrieved the \hi~image from The H I Nearby Galaxy Survey (THINGS; \cite{Walter2008}) with the Very Large Array (VLA). 
We used the natural weighted maps with a resolution of $15\farcs2\times11\farcs4$ and converted the integrated intensity to a surface density following \citet{Walter2008}.
The uncertainty of  \atom{H}{}{}\emissiontype{I} flux density was $\sim5\%$ due to the flux calibration.
The \hi~map was aligned and convolved to match the \ci~image.

\subsection{Infrared data}
The Very Nearby Galaxy Survey (VNGS, \cite{Bendo2012}) observed M83 
using Herschel/PACS at 70~\um~and 160~\um~and with Herschel/SPIRE at 250~\um, 350~\um, and 500~\um.  
The full width at half-maximum (FWHM) of the point spread functions were 
6\farcs0, 12\farcs0, 18\farcs2, 24\farcs5, and 36\farcs0 
for 70~\um,  160~\um, 250~\um,  350~\um, and 500~\um, respectively.
The calibration uncertainties corresponding to 70~\um,  160~\um, 250~\um,  350~\um, and 500~\um, were 0.03, 0.05, 0.07, 0.07, and 0.07, respectively \citep{Foyle2012}.
Using the images at each wavelength, \citet{Foyle2012} derived the dust temperature and mass surface density of M83 through the modified blackbody fitting in the IR regime on each pixel with 36\farcs0 resolution.
Consequently, they found that the temperature ranged from 20 to 30~K over the entire disk in both cases of a variable and constant dust emissivity index $\beta$.
We, therefore, used the images at 70~\um, 160~\um, and 250~\um~in this work (the peak of the blackbody curve is within the wavelengths)
\footnote{The image at 350\um~is also utilized only to effectively constrain the initial parameter for the modified blackbody fitting through the three wavelengths.
For the fitting including the 350\um~data, the images at 70\um, 160\um, and 250\um~were aligned and convolved to match the 350\um~images}, 
obtaining the properties on an angular scale matching with the spatial resolution of \ci.
The images were all aligned and convolved to the \C~image after applying calibration and conversion to Jy/sr$^{-1}$, following  \citet{Foyle2012}.

\begin{sidewaysfigure*}
\centering
 \begin{center}
  \includegraphics[width=\linewidth]{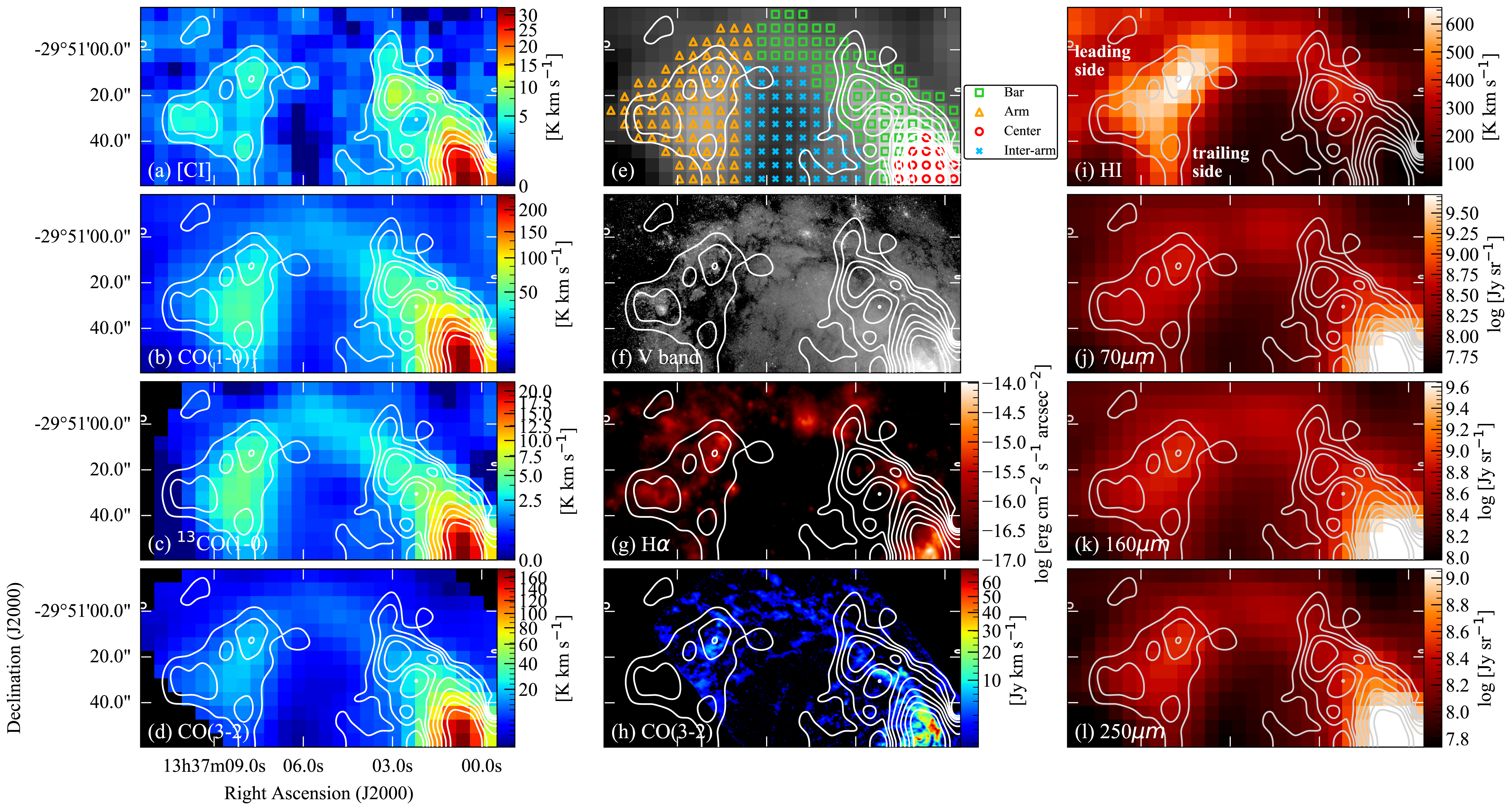}
 \end{center}
 \caption{ 
 Integrated intensity maps of 
 (a) \ci, 
 (b) \atom{CO}{}{12}(1--0) 
 (c) \atom{CO}{}{13}(1--0).
 (d) \atom{CO}{}{12}(3--2).
In panel~(e), the regions defined as central region, barred region, arm region, and inter-arm region are denoted by 
circles, squares, triangles, and crosses, respectively.
Panels~(f), (g), and (h) are V-band, \atom{H}{}{}\emissiontype{$\alpha$}, and \atom{CO}{}{12}(3--2) images at their original resolutions.
In panel~(d), only the TP array data is used for the \atom{CO}{}{12}(3--2) image, whereas the data of the 12-m, 7-m, and TP arrays are combined for the image in panel~(h).
Therefore, the image area is limited by the field of view of the12-m array.
Panels~(i), (j), (k), and (l) depict the integrated intensity map of \atom{H}{}{}\emissiontype{I} and surface brightness maps at 70~$\mu$m, 160~$\mu$m, and 250~$\mu$m, respectively.
Contours in all panels show the integrated intensity of \ci, corresponding to 3, 5, 7, 10, 15, 20, 25, and 30~$\sigma$, where 1$\sigma=$ 0.85~K~km~s$^{-1}$,
 }
 \label{fig:results}
\end{sidewaysfigure*}

\section{Results}
\label{sec:results}
\subsection{Distribution of \ci~intensity and comparison with data corresponding to other wavelengths}
Figure~\ref{fig:results}(a) shows the integrated intensity map ($I = \int T_{\rm mb}~dv$) of \ci, where $T_{\rm mb}$ is the main beam brightness temperature, 
and the contours of $I_{\rm \C(1-0)}$ are superposed on the remaining panels (Figure~\ref{fig:results}(b)--(l)).
A maximum integrated intensity of $I_{\rm \C(1-0)}=32.68\pm1.75$~K~km~s$^{-1}$ is found at the center, $(\alpha, \delta)_{\rm J2000.0} =$ (\timeform{13h37m00s.48},  \timeform{-29D51'56''.48}).
The integrated intensity maps of \co($J = 1-0$), \CO($J = 1-0$), and \co($J = 3-2$) with the same angular resolution as that of \ci~are shown in Figure~\ref{fig:results}(b)--(d). 
The uncertainty of the integrated intensity was calculated as $\Delta I = T_{\rm rms} \sqrt{\Delta V_{\rm int} V_{\rm ch}}$, 
where $T_{\rm rms}$ is the rms noise calculated over the emission-free channels, $\Delta V_{\rm int}$ is the full velocity width of the emission channels, and $V_{\rm ch}=10$~km~s$^{-1}$ is  the velocity width of a single channel.
The uncertainties in \ci, \co($J = 1-0$),\co($J = 3-2$), and \CO($J = 1-0$)
were typically 0.85, 0.69, 0.16, and 0.14~K~km~s$^{-1}$, respectively.
We defined the central region, bar region, spiral arm region, and the inter-arm region by referring to the \atom{CO}{}{}(1--0) map (Figure~\ref{fig:results}(e)).
A comparison of the distributions of \ci~and \atom{CO}{}{} lines in Figure~\ref{fig:results}(b)--(d) reveals that both are strong in the central region and are similarly distributed in the bar region, whereas \ci~is distributed outside the \atom{CO}{}{} gas in the spiral arm.

We compared the \ci~distribution with the V-band (obtained from the \ci~mapping area in Figure~\ref{fig:m83}), \atom{H}{}{}\emissiontype{$\alpha$}, and \atom{CO}{}{}(3--2) images at their original resolutions in Figure~\ref{fig:results}(f)--(h).
The continuum-subtracted  \atom{H}{}{}\emissiontype{$\alpha$} image was observed by the Survey for Ionization in Neutral Gas Galaxies (SINGG, \cite{Meurer2006}) using the Cerro Tololo Inter-American Observatory (CTIO) 1.5 m telescope. 
The data were acquired from the NASA/IPAC Extragalactic Database. 
The \atom{CO}{}{}(3--2) image with a high spatial resolution of $0\farcs7$ was produced using the ALMA 12-m, 7-m, and TP arrays in the same manner as described in section~\ref{sec:co}.
The \ci~distribution in the arm region, especially in the leading side (assuming trailing spiral arms), was in good agreement with the \atom{H}{}{}\emissiontype{$\alpha$} and \atom{CO}{}{}(3--2) distributions, although the \atom{CO}{}{}(3--2) image does not entirely cover the arm region due to the limitation of observing the area with the 12-m array.

Figures~\ref{fig:results}(i)--(l) compare the \ci~map with those of the integrated intensity of \hi~and the surface brightness of 70~\um, 160~\um, and 250~\um.
The \hi~ emission has a  hole at the central region, and the patterns of the \hi~and \ci~spiral arms are consistent not only on the leading side but also the trailing side.
In addition, the brightness of the image at 70~\um~is stronger on the leading side of the arm than on the trailing side, whereas that at 250~\um~is extended throughout the arm as with \atom{CO}{}{}(1--0).
This is compatible with the result that \atom{CO}{}{}(1--0) and 100~\um~are not well correlated \citep{Crosthwaite2002}.
Based on the comparison of the dust color temperatures for the 70, 160, 250, 350, and 500~\um~bands with \atom{H}{}{}\emissiontype{$\alpha$}, 
\citet{Bendo2012} argued that the dust emission at wavelengths shorter than 160~\um~could be affected by the star-forming region, whereas the wavelengths longer than 250~\um~could be less affected and trace cold dust.
Furthermore, the fact that \hi~correlates better with 70~\um~and \atom{H}{}{}\emissiontype{$\alpha$} than with 250~\um~and that it is strong on the leading side of the arm,
is consistent with the understanding that \hi~ is a photodissociation product (e.g., \cite{Rand1992}, \cite{Hirota2018}).

\subsection{\C~-- \atom{CO}{}{} correlation}
\begin{figure*}[tph]
 \begin{center}
  \includegraphics[width=\linewidth]{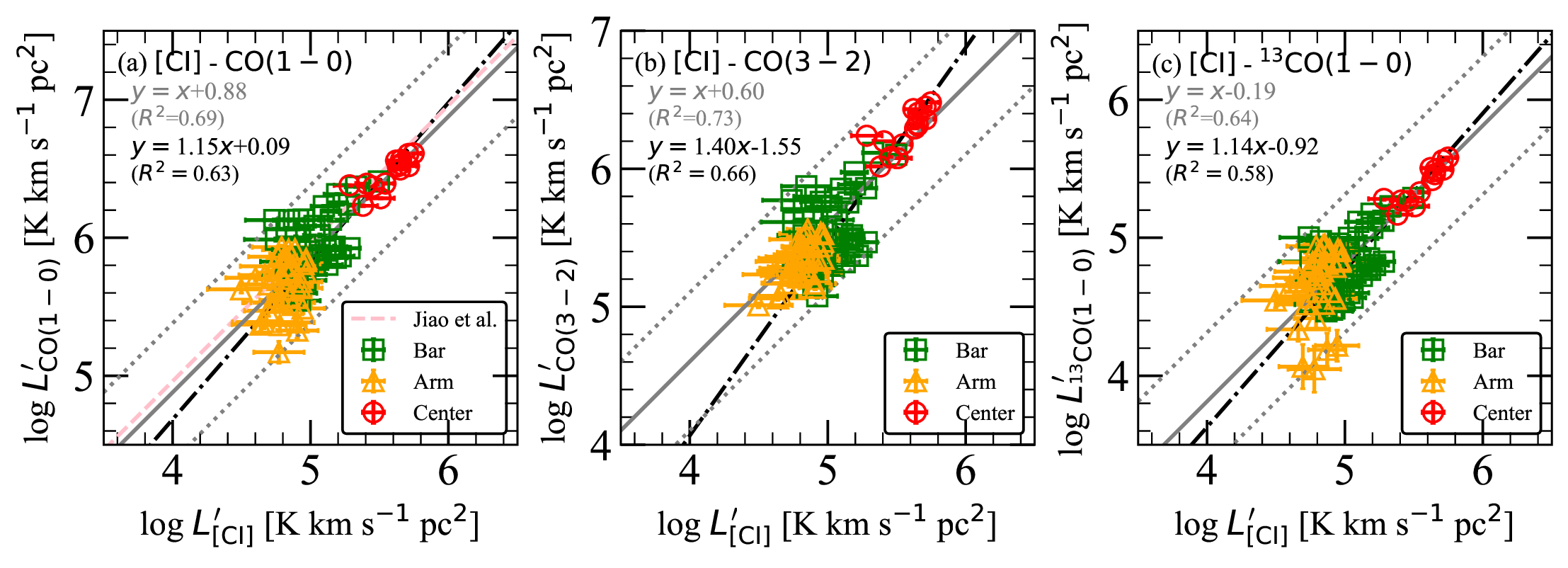}
 \end{center}
 \caption{
 Plots of the line luminosities of (a) \atom{CO}{}{}(1--0), (b) \atom{CO}{}{}(3--2), and (c) \atom{CO}{}{13}(1--0) against that of \ci~above 4$\sigma$.
 The circle, square, and triangle symbols correspond to the central region, barred region, and arm region, respectively, corresponding to those in Figure~\ref{fig:results}(e). 
 The best-fit relation with a fixed slope of unity for all regions is shown by the gray solid line, and the best-fit relation without fixing the slope 
 is shown by the black dashed-dotted line.
 The dotted lines represent 0.5~dex offset from the solid line. 
 The (pink) dashed line in panel~(a) is the best-fit relation with a fixed slope of unity for nearby galaxies \citep{Jiao2019}. 
}
 \label{fig:correlation}
\end{figure*}
\begin{table*}
  \tbl{Determined Parameters and Correlation Coefficient 
  }{%
  \begin{tabular}{ccccccc}
      \hline
      Line 		& Region & $A$ 		& $B$ 		& $R_{A,B}^2$	& $C$ 		& $R_C^2$ \\ 
      \hline
      \atom{CO}{}{}(1--0) 	& All 		&$1.15\pm0.05$	&$0.09\pm0.26$	& $0.63$		&$0.88\pm0.01$	&$0.69$\\
      ~  		& Center	&$0.65\pm0.08$	&$2.83\pm0.45$	& $0.79$		&$0.92\pm0.02$	&$0.57$\\      
      \atom{CO}{}{}(3--2) 	& All 		&$1.40\pm0.06$	&$-1.55\pm0.30$	& $0.66$		&$0.60\pm0.02$	&$0.73$\\
      ~  		& Center	&$0.93\pm0.12$	&$1.11\pm0.65$	& $0.78$		&$0.70\pm0.02$	&$0.78$\\      
      \atom{CO}{}{13}(1--0) 	& All 		& $1.14\pm0.05$	&$-0.92\pm0.28$	& 0.58		&$ -0.19\pm0.02$	& 0.64\\
      ~  		& Center	&$0.82\pm0.07$	&$0.85\pm0.41$	& 0.88		&$-0.16\pm0.02$	& 0.83\\
      \hline
    \end{tabular}}\label{tab:fit}
\end{table*}

We compared the line luminosities between \atom{CO}{}{}(1--0) and \ci~in a pixel-by-pixel manner, and both the lines were detected above the $4\sigma$ noise level (Figure~\ref{fig:correlation}(a)).
The line luminosity, $L'_{\rm line}$, in units of K~km~s$^{-1}$~pc$^2$ was calculated as the product of the integrated intensity, $I_{\rm line}$, and the projected pixel area, $A_0$, in pc$^2$, as: $L'_{\rm line} = I_{\rm line} A_0$ (cf. \cite{Solomon2005}).
Although the correlation between  \atom{CO}{}{}(1--0) and \ci~is worse in the arm region in Figure~\ref{fig:results} and \ref{fig:correlation}, 
the overall distribution of both luminosities in the whole disk is correlated with each other in Figure~\ref{fig:results}.
The  \atom{CO}{}{}(1--0)--\ci~relation was fitted by: 
\begin{eqnarray}
\log L'_{\rm \atom{CO}{}{}(1-0)} &=& A\log L'_{\rm \C(1-0)} +B~[{\rm K~km~s^{-1}~pc^{2}}], \\
L'_{\rm \atom{CO}{}{}(1-0)} &=& 10^B~{L'}_{\rm \C(1-0)}^A ,
\end{eqnarray}
using orthogonal distance regression (ODR).
The coefficients $(A, B)$ were derived as $(1.15\pm0.05, 0.09\pm0.26)$ 
with the coefficient of determination, $R^2=0.63$ (black dashed-dotted line in Figure~\ref{fig:correlation}(a), Table~\ref{tab:fit}).
The slope was slightly steeper than that of the relation for nearby galaxies that were observed at a scale of $\sim1$~kpc, as shown by \citet{Jiao2019}, 
where reportedly $(A, B) = (1.04 \pm 0.02, 0.74\pm0.12)$.
The steep  slope is mainly caused by the relation in the arm  region (triangle symbols in Figure~\ref{fig:correlation}), 
lying below the relation in the central region.
The least square fit to the data only in the central region yields coefficients as $(A, B)=(0.65\pm0.08,  2.83\pm0.45)$, 
wherein the observed shallow slope is in  agreement with that  for the central region of the nearby starburst/Seyfert galaxy NGC~1808 ($0.686\pm0.010$, \cite{Salak2019}) and for LIRG IRAS F18293-3413 ($0.65\pm0.01$, \cite{Saito2020}).

By fitting the relationship of the line luminosities in all regions, including the center, bar, and  arm, with a fixed slope of unity using ODR, 
i.e., 
\begin{eqnarray}
\log L'_{\rm \atom{CO}{}{}(1-0)} &=& \log L'_{\rm \C(1-0)} +C~[{\rm K~km~s^{-1}~pc^{2}}], \\
L'_{\rm \atom{CO}{}{}(1-0)} &=& 10^C~L'_{\rm \C(1-0)}, 
\label{eq:C}
\end{eqnarray}
the coefficient  $C=0.88\pm0.01$  was derived with $R^2=0.69$. 
This is represented by the gray solid line in Figure~\ref{fig:correlation}, 
where the intercept $C$ corresponds to the logarithm of the luminosity ratio of \atom{CO}{}{}(1--0) to \ci.
The intercept of the data only in the central region, $C=0.92\pm0.02$ with $R^2=0.57$, is comparable to that in nearby galaxies $(C=0.96\pm0.01)$, as reported by \citet{Jiao2019}.

Thus far, the \ci~and \atom{CO}{}{}(1--0) luminosity relation in  galaxies has been determined  by measurements of \atom{CO}{}{}-bright sources, such as 
the galactic centers and (U)LIRGs.
The relationship in the 
\atom{CO}{}{}-bright region in the arm of M~83 follows the main relation ($C\sim1$),  corresponding to 
the integrated intensity ratio, $I_{\rm \C}/I_{\rm \atom{CO}{}{}(1-0)}$ [denoted by \Rcico $(=L'_{\rm \C}/L'_{\rm \atom{CO}{}{}(1-0)})$], of $\sim0.1$ (equation~(\ref{eq:C})), 
which is comparable to that for the central region of 30 nearby galaxies, \Rcico~$=0.16\pm0.08$ \citep{Israel2020}.
However,  the data from the leading side of the arm (i.e., \atom{CO}{}{}-dark region) is deviated 
from the main relation, i.e., \ci~enhancement to \atom{CO}{}{}(1--0).
\citet{Ojha2001} found that 
 \Rcico~at the Galactic center $(\sim 0.1)$ is lower than that in the disk $(\sim 0.3-0.4)$.
Even for nearby galaxies, the similar trend that  \Rcico at the center is lower than that in the disk has also been reported by \citet{Gerin2000}.
To understand the general relation between the [\atom{C}{}{}\emissiontype{I}](1--0) and \atom{CO}{}{}(1--0) luminosities in galaxies, 
their measurements, both in the central as well disk regions, are necessary.

Figures~\ref{fig:correlation}(b) and (c) compare 
the line luminosities of \atom{CO}{}{}(3--2) to~\ci~and \atom{CO}{}{13}(1--0) to \ci, respectively, on a pixel-by-pixel basis.
The relations were fitted in the same manner as that of [\atom{C}{}{}\emissiontype{I}](1--0)--\atom{CO}{}{}(1--0) luminosities, and the resultant coefficients are summarized  in Table~\ref{tab:fit}. 
Using the central region data, 
nearly linear fits were obtained for the relations of [\atom{C}{}{}\emissiontype{I}](1--0)--\atom{CO}{}{}(3--2) and [\atom{C}{}{}\emissiontype{I}](1--0)-$^{13}$\atom{CO}{}{}(1--0).
The slope, $A$,  for [\atom{C}{}{}\emissiontype{I}](1--0)--\atom{CO}{}{}(3--2)  is in good agreement with that obtained by \citet{Salak2019} for the  nearby galaxy NGC~1808 ($A=0.942\pm0.014$).
As with the relation for the  [\atom{C}{}{}\emissiontype{I}](1-0)--\atom{CO}{}{}(1-0) relation, 
the slope, $A$, of [\atom{C}{}{}\emissiontype{I}](1--0)--\atom{CO}{}{}(3--2) and [\atom{C}{}{}\emissiontype{I}](1--0)-\atom{CO}{}{13}(1--0), 
calculated by using the data from all the regions, is steeper than that in the central region, 
as the arm region is located below the relation in the central region. 

We also fitted the [\atom{C}{}{}\emissiontype{I}](1--0)--\atom{CO}{}{}(3--2) and [\atom{C}{}{}\emissiontype{I}](1--0)-\atom{CO}{}{13}(1--0) luminosities with a fixed slope of unity. 
The calculated intercepts, $0.70\pm0.02$ for  [\atom{C}{}{}\emissiontype{I}](1--0)--\atom{CO}{}{}(3--2) and $-0.19\pm0.02$ for  [\atom{C}{}{}\emissiontype{I}](1--0)-\atom{CO}{}{13}(1--0), were compatible with the integrated intensity ratios, $I_{\rm \C(1-0)}/I_{\rm \atom{CO}{}{}(3-2)}$ ($\sim0.1-0.3$) 
and $I_{\rm \C(1-0)}/I_{\rm \atom{CO}{}{13}(1-0)}$ ($\sim 1-10$ ) for the central region of the nearby galaxy NGC~613 \citep{Miyamoto2018}.
The  large scatter ($\pm0.5$~dex) in the disk region data for both relations may reflect 
the physical properties of the gas, 
e.g., kinetic temperature for  [\atom{C}{}{}\emissiontype{I}](1--0)--\atom{CO}{}{}(3--2) relation \citep{Ikeda1999} and  optical depth for   [\atom{C}{}{}\emissiontype{I}](1--0)-\atom{CO}{}{13}(1--0) relation \citep{Miyamoto2018}.

\subsection{Dust temperature and dust mass surface density}
\label{sec:dust}
\begin{figure}[th]
 \begin{center}
 \includegraphics[width=\linewidth]{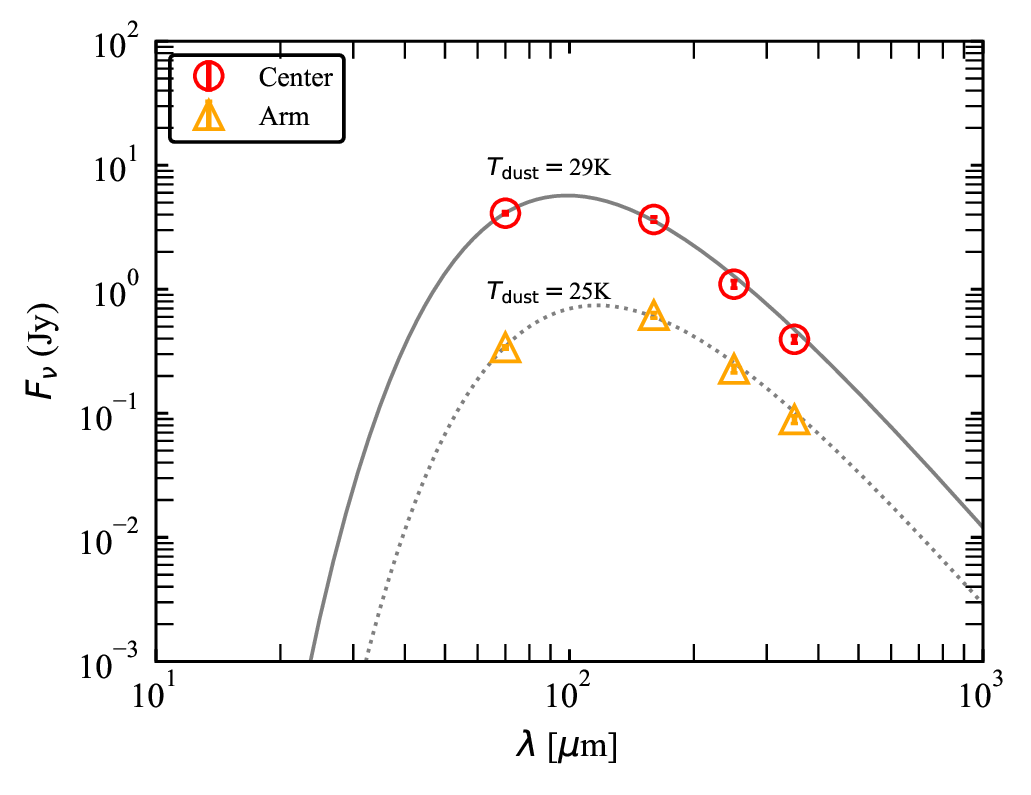}
 \end{center}
 \caption{ 
Plots for two examples of the initial 
modified blackbody
fitting at 70, 160, 250, and 350~$\mu$m for
pixels at the central region and arm regions, denoted by circle and triangle symbols, respectively, in Figure~\ref{fig:Dust}.
The best-fitting  is shown as a solid curve (center) and a dotted curve (arm). 
The best-fitting parameters obtained by using the data at 350~\um~are adopted as the initial values for the modified blackbody fitting at each pixel using the data only at 70~$\mu$m, 160~$\mu$m, and 250~$\mu$m.
}
 \label{fig:sed}
\end{figure}

\begin{figure}[th]
 \begin{center}
 \includegraphics[width=\linewidth]{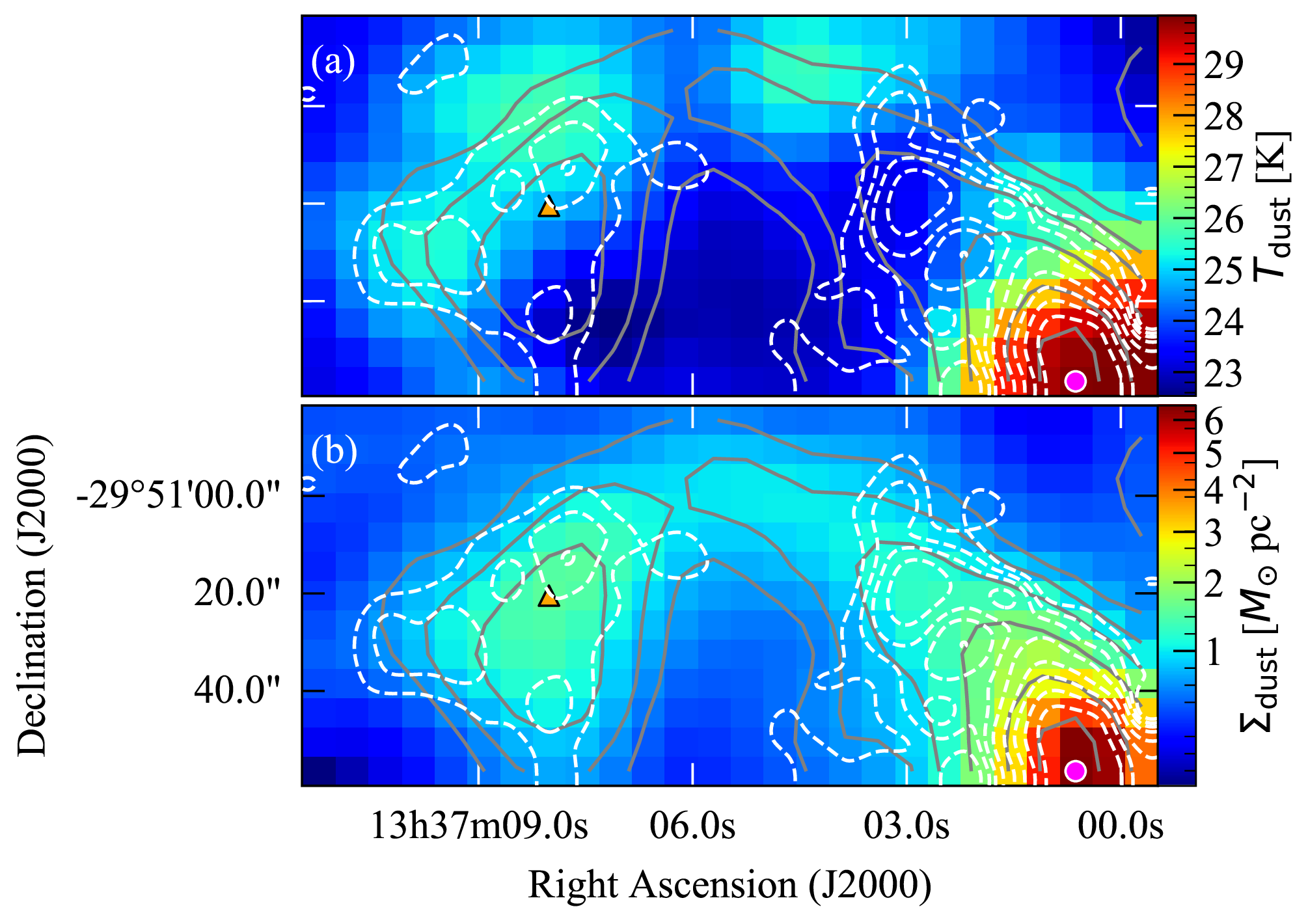}
 \end{center}
 \caption{ 
Spatial distribution (color) of (a) dust temperature, $T_{\rm dust}$, and (b) dust mass surface density, $\Sigma_{\rm dust}$, overlaid with 
the integrated intensity of \ci~(white dashed line) and \atom{CO}{}{}(1--0) (gray solid line).
The contours of \ci~are the same as in Figure~\ref{fig:results} and the contours of \atom{CO}{}{}(1--0) are 
20, 30, 50, 100, 200, and 300~$\sigma$, where 1$\sigma=$ 0.686~K~km~s$^{-1}$.
The circle at the center and triangle in the arm indicate the location of the pixels for which the best-fitting plots are shown in Figure~\ref{fig:sed}.
}
 \label{fig:Dust}
\end{figure}

\begin{figure}[th]
 \begin{center}
 \includegraphics[width=\linewidth]{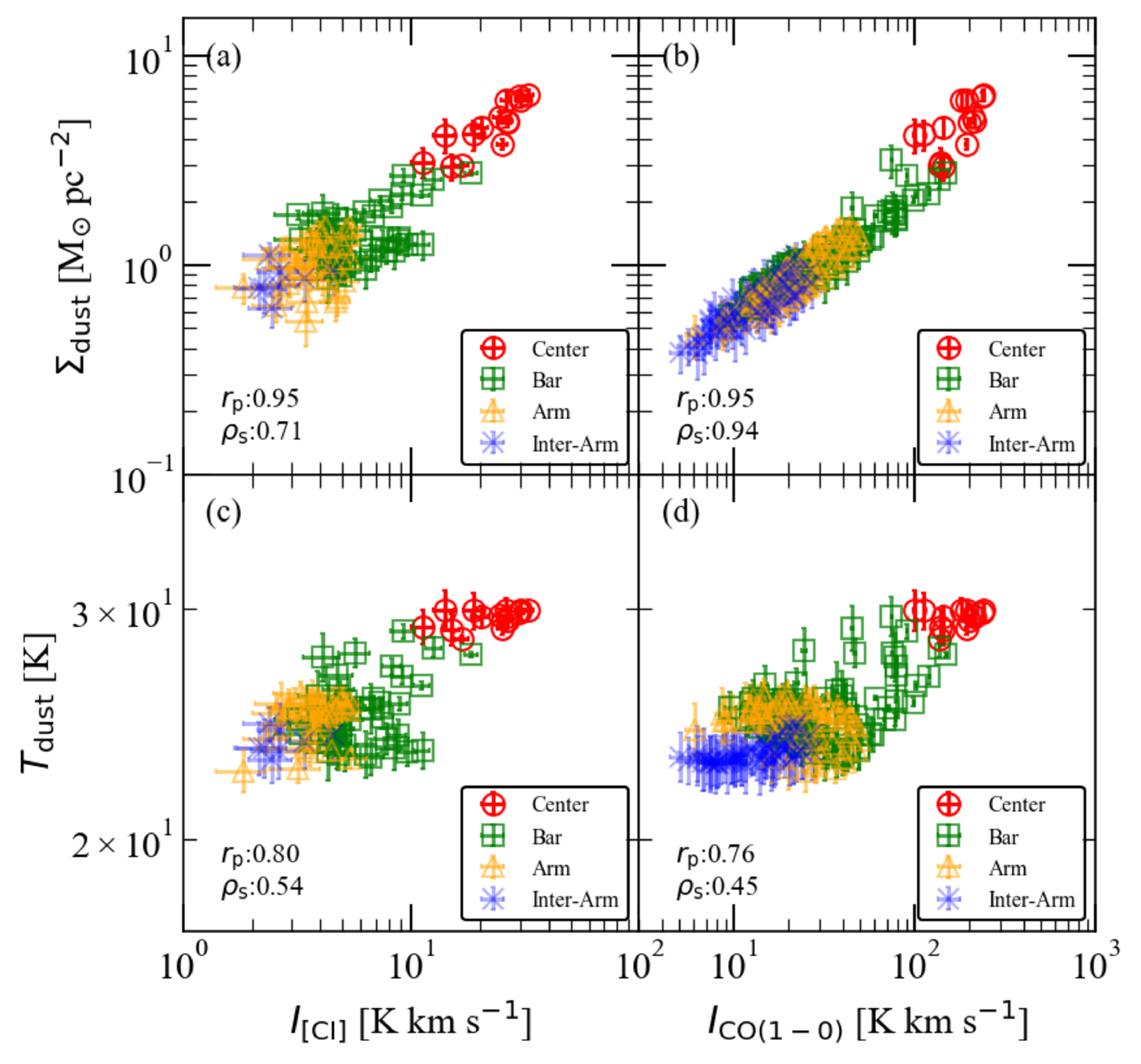}
 \end{center}
 \caption{ 
The  correlations of $\Sigma_{\rm dust}$ to (a) $I_{\rm \C(1-0)}$ and (b) $I_{\rm \atom{CO}{}{}(1-0)}$, and of $T_{\rm dust}$ to (c) $I_{\rm \C(1-0)}$ and (d) $I_{\rm \atom{CO}{}{}(1-0)}$.
The circle, square, triangle, and cross symbols indicate the central region, barred region, arm region, and inter-arm region, respectively.
Pearson’s correlation coefficient ($r_{\rm p}$) and Spearman’s rank correlation coefficient ($\rho_{\rm s}$) are labeled in each of the panels.
}
 \label{fig:Dust-CICO}
\end{figure}

To determine the dust temperatures of M83, we executed 
a single-temperature modified blackbody fitting 
in the far-IR regime   
at each pixel as:
\begin{eqnarray}
I({\rm \nu},T_{\rm dust}) = N \nu^{\beta} B({\rm \nu},T_{\rm dust}), \
\label{eq:SED}
\end{eqnarray}
where $I({\rm \nu},T_{\rm dust})$ is the flux density at frequency $\nu$ and dust temperature $T_{\rm dust}$, 
$B({\rm \nu},T_{\rm dust})$  the Planck function, 
$N$ a constant related to the column density for matching the model to the observed fluxes, 
and $\beta$  the dust emissivity index.
The parameter $\beta$  
originates from the dust opacity function, $\kappa_{\rm \nu} = \kappa_0(\nu/\nu_0)^{\beta}$.
\citet{Foyle2012} examined the spatial distribution of $\beta$ and dust temperature in M83 with a spatial resolution of \timeform{36"}  through the modified blackbody fitting
on the data of  five IR wavelengths (70, 160, 250, 350, and 500~\um).
They found that $\beta$ is close to 2 over much of the galaxy, and that the resulting dust temperature in the inner region  ($r<\timeform{3'}$) is between 20~K and 30~K  in both cases of variable  $\beta$ and constant $\beta(=2)$.
Therefore, we used the constant $\beta(=2)$ to our analysis, and fit equation~(\ref{eq:SED}) to the flux measurements at 70~\um, 160~\um, and 250~\um~to determine the best-fitting temperature, $T_{\rm dust}$, and constant $N$ at each pixel.
For the fitting, we adopted the initial values of the parameters that were 
derived from the modified blackbody fitting on the data at four wavelengths, including the data for 350~\um, where all data were convolved  to the 350~\um~resolution (24\farcs5) (Figure~\ref{fig:sed}).

The dust optical depth, $\tau_{\rm dust}$, can be calculated as the ratio of 
the measured flux density 
 to the Planck function at a certain dust temperature and wavelength \citep{Planck2011} as: 
\begin{eqnarray}
\tau_{\rm dust}(\lambda) = \frac{I(\nu)}{B({\rm \nu},T_{\rm dust})}.
\label{eq:tau}
\end{eqnarray}
Using the results from our fitting, we calculated the optical depth at 250~\um~and found that it was the highest at the center $\tau_{\rm dust}\sim0.09$ and its mean value over the galaxy was $\tau_{\rm dust}=0.005$.
Accordingly, the assumption of  optically thin emission at 250~\um~in the galaxy is reasonable.

The dust mass at a specific frequency, $\nu$, is calculated by the following equation: 
\begin{eqnarray}
M_{\rm dust}=\frac{S_{\rm \nu}D^2}{\kappa_{\rm \nu} B{(\rm \nu, T_{\rm dust}})},
\label{eq:dmass}
\end{eqnarray}
where
$S_{\rm \nu}$ is the flux density at a certain wavelength from our modified blackbody fit,
$D$  the distance to M83, $\kappa_{\rm \nu}$  the dust opacity, and $B({\rm \nu, T_{\rm dust}})$  the Planck function.
Here, we adopted a value of 0.398~m$^2$~kg$^{-1}$  for the dust opacity at 250~\um, denoted as $\kappa_{250}$, from \citet{Draine2003}.
Using the results from our fitting, we calculated the dust mass at 250~\um~and then the dust mass surface density. 
However, notably, $\kappa_{\rm \nu}$ is a factor combining various dust grain properties (e.g., size distributions, morphology, density, and chemical composition); hence, it depends on the environment, such as interstellar medium (ISM) density. 
The size of the dust grains in the dense  ISM are predicted to be larger, due to the coagulation of grains, and  larger grains should show enhanced emissivity (e.g., \cite{Kohler2012}).
In contrast, \citet{Clark2019} found that $\kappa_{\rm 500}$ is negatively correlated with the ISM density and 
varies by a factor of 5.5 in the entire disk of M~83.
Thus, although more accurate measurements of $\kappa_{\rm \nu}$ are required, 
its variance at 500~$\mu$m in the studied area would introduce an uncertainty by a factor of $\sim 2$ to $M_{\rm dust}$,   
assuming the similar variation of $\kappa_{\rm 250}$  with $\kappa_{\rm 500}$, 
and consequently, an uncertainty by much less than a factor of 2 to $\mathrm{GDR}$ and $\alpha_{\rm \atom{CO}{}{}}$ in the self-consistent calculations  (equation (\ref{eq:gas})).

Figure~\ref{fig:Dust} shows the distributions of the dust temperature, $T_{\rm dust}$,  and dust mass surface density, $\Sigma_{\rm dust}$,  with contours of the integrated intensities of \ci ~and \atom{CO}{}{}(1--0).
The temperature is the highest at the center ($\sim30$~K), 
whereas those in the bar and arm regions  are offset to the leading edge of the \atom{CO}{}{} and dust distributions, which can be caused by the high star formation activity,  as seen in the \atom{H}{}{}\emissiontype{$\alpha$} distribution (Figure~\ref{fig:results}(g)).
The spatial distribution of $\Sigma_{\rm dust}$ is in good agreement with the \atom{CO}{}{}(1--0) distribution.
The dust temperature, dust mass surface density, and the offsets of their distributions are  consistent with the results reported by \citet{Foyle2012}.

Figure~\ref{fig:Dust-CICO} shows the plot of $\Sigma_{\rm dust}$ and $T_{\rm dust}$ as functions of the integrated intensities of ~\ci~and \atom{CO}{}{}(1--0), 
with Pearson's correlation coefficient ($r_{\rm p}$) as a parametric measure and Spearman's rank correlation coefficient ($\rho_{\rm s}$) as a non-parametric measure. 
The relations of $\Sigma_{\rm dust}$ to both $I_{\rm \C(1-0)}$ and $I_{\rm \atom{CO}{}{}(1-0)}$ show high correlation ($\rho_{\rm s}=0.71, 0.94$).
Contrarily, $I_{\rm \C(1-0)}$ and $I_{\rm \atom{CO}{}{}(1-0)}$ are moderately correlated with $T_{\rm dust}$ ($\rho_{\rm s}=0.54, 0.45$).
The slightly higher correlation coefficient ($\rho_{\rm s}=0.54$) for  the  $I_{\rm \C(1-0)}-T_{\rm dust}$ relation than that of $I_{\rm \atom{CO}{}{}(1-0)}-T_{\rm dust}$ ($\rho_{\rm s}=0.45$) would imply that \C~is more sensitive to the temperature than \atom{CO}{}{}(1--0).
Additionally, the high correlation coefficients for the relation between $\Sigma_{\rm dust}$ and $I_{\rm \atom{CO}{}{}(1-0)}$ and their consistent distributions  
indicate that the cold dust is well mixed with the molecular gas.

\section{Discussion}
\label{sec:discussion}
\subsection{Derivation of the gas-to-dust ratio and the \atom{CO}{}{}-to-\atom{H}{}{}$_2$ conversion factor, $\alpha_{\rm \atom{CO}{}{}}$}
\label{subsec:alpha_co}
\begin{figure}[th]
 \begin{center}
 \includegraphics[width=0.8\linewidth]{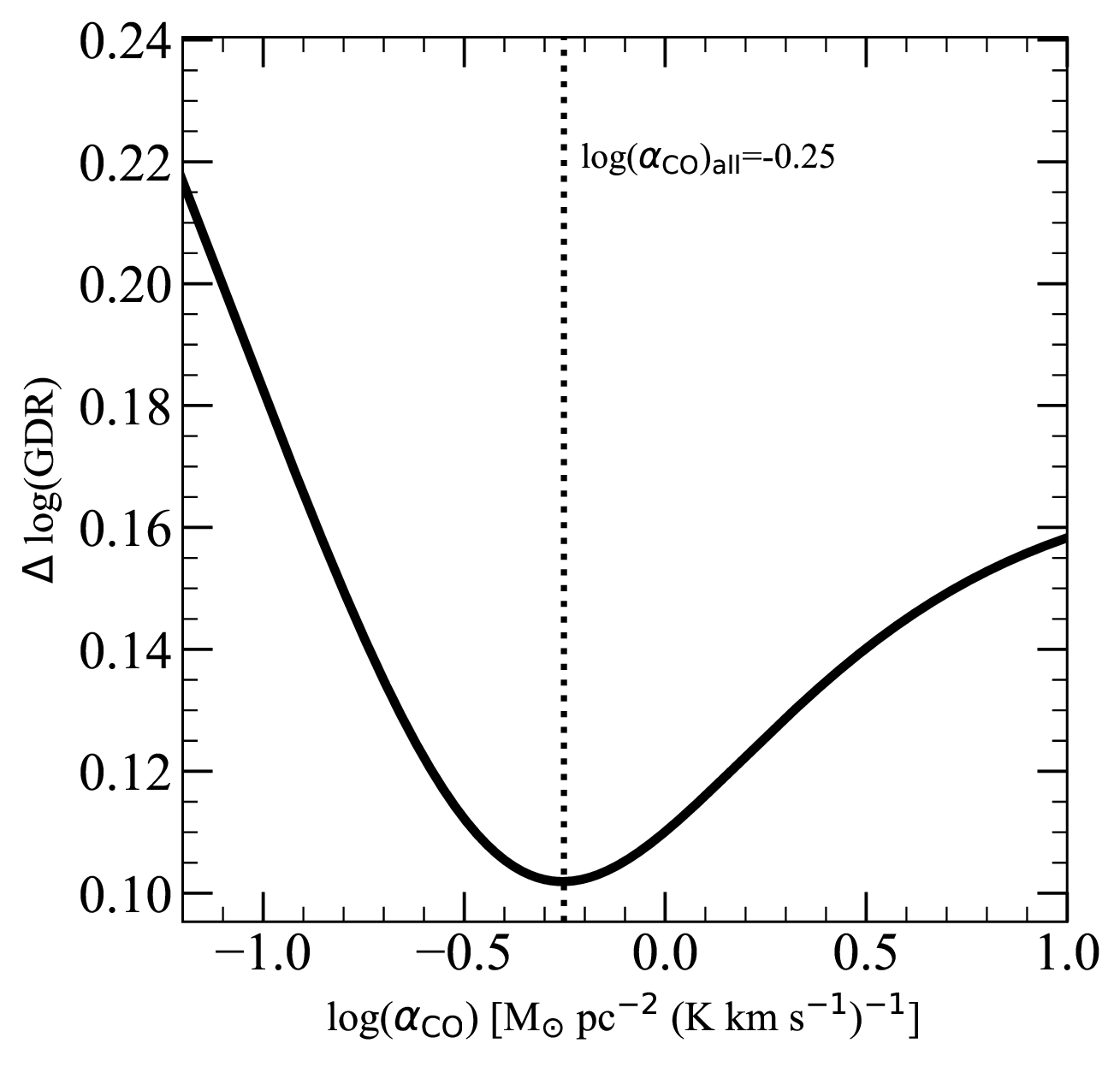}
 \end{center}
 \caption{ 
Scatter in the $\log(\mathrm{GDR})$ as a function of $\alpha_{\rm \atom{CO}{}{}}$.
The $\log(\mathrm{GDR})$ scatter in our mapping area is minimized when the best-fit $\log(\alpha_{\rm \atom{CO}{}{}}) = 0.25$ is employed.
}
 \label{fig:alpha}
\end{figure}

Given that dust and gas are well mixed and are linearly correlated, 
the total gas mass surface density $\Sigma_{\rm gas}$ can be estimated from the dust surface density, $\Sigma_{\rm dust}$  (see equation~(\ref{eq:gas})).
The molecular gas surface density, $\Sigma_{\rm mol}$, can be measured by 
including the contribution of the atomic surface density, $\Sigma_{\rm atom}$, 
which is derived from  \hi, 
to the total gas surface density 
$\Sigma_{\rm gas} (= \Sigma_{\rm atom} + \Sigma_{\rm mol})$.
Conversely,  $\Sigma_{\rm mol}$ is derived from \atom{CO}{}{} by using a constant conversion factor,  $\alpha_{\rm \atom{CO}{}{}}$, in which a factor of 1.36 is included to account for helium.
\citet{Leroy2011} proposed a technique to solve  $\mathrm{GDR}$ and $\alpha_{\rm \atom{CO}{}{}}$ simultaneously in Local Group galaxies  by applying equation~(\ref{eq:gas}) to the maps of \atom{CO}{}{}, \hi, and IR emission under the assumptions that the $\mathrm{GDR}$ is constant over a region of galaxies
and also 
does not vary between the atomic and molecular phases.
The best $\alpha_{\rm \atom{CO}{}{}}$ in each target region can be determined by  minimizing the scatter in $\mathrm{GDR}$ as a function of the  $\alpha_{\rm \atom{CO}{}{}}$; moreover, the $\mathrm{GDR}$ is also determined simultaneously.
Refining the technique, \citet{Sandstrom2013} derived the distribution of $\mathrm{GDR}$ and $\alpha_{\rm \atom{CO}{}{}}$ in nearby star-forming galaxies on the kpc scales, and showed $\mathrm{GDR}$ as an approximately linear function of the metallicity, with a slope of $\sim-0.85$, unlike $\alpha_{\rm \atom{CO}{}{}}$.
Considering a metallicity gradient with a radius  in the inner disk region of M83 ($<0.5R_{\rm 25}$) and with a slope of $-0.303~{\rm dex}~R_{25}^{-1}$ \citep{Hernandez2019}, 
it can establish that the assumption that the $\mathrm{GDR}$  is constant across the area studied in this work is reasonable.

Notably, the limitations of the method, 
wherein contamination from dust mixed with invisible components, such as opaque \atom{H}{}{}\emissiontype{I} and ionized gas,  are considered as negligible. 
However, \citet{Israel2001} suggested that approximately half of all hydrogen  is associated with ionized carbon in the central region of M~83. 
If a large amount of photodissociation products of \atom{CO}{}{}, which is associated with the amount of hydrogen exists even in the disk region, then $\alpha_{\rm \atom{CO}{}{}}$ is overestimated by a factor of approximately 2.

$\mathrm{GDR}$ and dust opacity, $\kappa_{\rm \nu}$, may change  within galaxies due to metallicity and gas density (e.g., \cite{Draine2007}, \cite{Remy2014}, \cite{Galliano2018} and references therein).
Considering a metallicity gradient in the studied area,  
the effect of the metallicity on GDR is expected to be minor in this work.
However, as described in section~\ref{sec:dust}, the variation in $\kappa_{\rm \nu}$ would introduce an uncertainly by less than a factor of 2 in $\mathrm{GDR}$ and $\alpha_{\rm \atom{CO}{}{}}$. 
In addition, uncertainties in $\mathrm{GDR}$ and $\alpha_{\rm \atom{CO}{}{}}$ can be attributed to the differences in dust emissivity between the molecular and atomic phases  (e.g., a factor of $\sim2$ for the Milky Way; \cite{Planck2011b}); however, 
this effect can be neglected in this work because both phases in M83 exhibit similar emissivity, as reported  by \citet{Clark2019}.
In conclusion, in this study, 
the derived $\mathrm{GDR}$ and $\alpha_{\rm \atom{CO}{}{}}$ have inherent uncertainties of at least a factor of 2.

Applying the same technique as \citet{Leroy2011}, we searched for  $\alpha_{\rm \atom{CO}{}{}}$ with the lowest scatter of the GDR in the galaxy, in the range of $\alpha_{\rm \atom{CO}{}{}}=0.05-50$~$M_{\Sol}$~pc$^{-2}$~(K~km~s$^{-1}$)$^{-1}$ with a step of 0.01~$M_{\Sol}$~pc$^{-2}$~(K~km~s$^{-1}$)$^{-1}$.
At each value of $\alpha_{\rm \atom{CO}{}{}}$, the  $\mathrm{GDRs}$ were calculated at all points in the \ci~mapping region. 
After dividing the $\mathrm{GDR}$ at each pixel  by the mean $\mathrm{GDR}$ in the region,  the scatter of the resulting values ($\Delta \log(\mathrm{DGR})$) was measured.
Figure~\ref{fig:alpha} presents the scatter in logarithm as a function of $\alpha_{\rm \atom{CO}{}{}}$.
The minimum $\mathrm{GDR}$ scatter can be found at $\alpha_{\rm \atom{CO}{}{}}\approx0.56$~$M_{\Sol}$~pc$^{-2}$~(K~km~s$^{-1}$)$^{-1}$ [$X_{\rm \atom{CO}{}{}}=0.26
\times10^{20}$~cm$^{-2}$(K~km~s$^{-1}$)$^{-1}$], 
corresponding to $\mathrm{GDR}\approx20$. 
These values are smaller than the averaged values for nearby galaxies, $\alpha_{\rm \atom{CO}{}{}}\approx$3.1~$M_{\Sol}$~pc$^{-2}$~(K~km~s$^{-1}$)$^{-1}$ and $\mathrm{GDR}\approx72$, reported by \citet{Sandstrom2013}. 
Applying the standard  $\alpha_{\rm \atom{CO}{}{}}(=3.1~M_{\Sol}~{\rm pc}^{-2}~({\rm K~km~s}^{-1})^{-1}
)$ to our datasets yields 
the $\mathrm{GDR}~\approx110$ in the area studied ($r\leq\timeform{160"}$), 
 which is compatible with the result ($\mathrm{GDR}\sim70-180$) reported by \citet{Foyle2012}. 
It has been reported that the conversion factor in the central region of M83, derived from the radiative transfer model using multiple \atom{CO}{}{} and \atom{CO}{}{13} lines, becomes  $X_{\rm \atom{CO}{}{}}=0.25\times10^{20}$~cm$^{-2}$(K~km~s$^{-1}$)$^{-1}$, corresponding to $\alpha_{\rm \atom{CO}{}{}}\sim$0.5~$M_{\Sol}$~pc$^{-2}$~(K~km~s$^{-1}$)$^{-1}$ \citep{Israel2001}, which is consistent with our result.
Nevertheless, 
in addition to the uncertainties as described before, 
the single temperature modified blackbody fitting that we used could lead to an overestimation of the dust mass under the high temperature environment, thereby yielding a low $\mathrm{GDR}$.

For applying the best fit $\alpha_{\rm \atom{CO}{}{}}$ and $\mathrm{GDR}$ values, 
the corresponding uncertainties in $\alpha_{\rm \atom{CO}{}{}}$ and $\mathrm{GDR}$ are estimated from the standard deviation of the values in the area 
and the systematic uncertainties arising from the uncertainty of the measured $\Sigma_{\hi}$, $I_{\rm \atom{CO}{}{}}$, and $\Sigma_{\rm dust}$  
for applying the best fit $\alpha_{\rm \atom{CO}{}{}}$ and $\mathrm{GDR}$ values.
The standard deviations become  0.26~$M_{\Sol}$~pc$^{-2}$~(K~km~s$^{-1})^{-1}$ in $\alpha_{\rm \atom{CO}{}{}}$ and 4.54 in $\mathrm{GDR}$. 
The systematic uncertainties are calculated  through a Monte Carlo test on the solutions by adding random noise to the measured $\Sigma_{\hi}$, $I_{\rm \atom{CO}{}{}}$, and $\Sigma_{\rm dust}$ according to the errors at each pixel.
The calculation was repeated 100 times using the randomly perturbed data values and the standard deviations of the results were found to be 0.07~$M_{\Sol}$~pc$^{-2}$~(K~km~s$^{-1})^{-1}$ for $\alpha_{\rm \atom{CO}{}{}}$ and 1.65 for the $\mathrm{GDR}$. 
Summing the uncertainties yields 0.27~$M_{\Sol}$~pc$^{-2}$~(K~km~s$^{-1})^{-1}$ for $\alpha_{\rm \atom{CO}{}{}}$ and 4.83 for the $\mathrm{GDR}$. 

Further, we  adopted other techniques to search for the best $\alpha_{\rm \atom{CO}{}{}}$ and $\mathrm{GDR}$, e.g., {\it ''Minimum fractional scatter in the $\mathrm{GDR}$''} and {\it ''Minimum $\chi^2$ of best-fit plane to $I_{\rm \atom{CO}{}{}}$, $N_{\rm HI}$, and $\Sigma_{\rm D}$''}, introduced by \citet{Sandstrom2013}.
In addition to that, we searched for the best $\alpha_{\rm \atom{CO}{}{}}$ and $\mathrm{GDR}$ on the plane so as to minimize the summation of $(\Sigma_{{\rm gas, (dust)}, i} - \Sigma_{{\rm gas, (atom+mol)}, i})^2/(\delta\Sigma_{\rm gas,(dust)}^2+\delta\Sigma_{\rm gas, (atom+mol)}^2)$, where 
$\Sigma_{{\rm gas, (dust)}, i} = \mathrm{GDR}~\Sigma_{{\rm dust}, i}$, 
$ \Sigma_{{\rm gas, (atom+mol)}, i} = \Sigma_{{\rm atom}, i}+\alpha_{\rm \atom{CO}{}{}}~I_{{\rm \atom{CO}{}{}}, i}$ at each pixel $i$, and $\delta\Sigma$ is the standard deviation in the region.
The results are in a range of 0.5--0.65~$M_{\Sol}$~pc$^{-2}$~(K~km~s$^{-1})^{-1}$ for $\alpha_{\rm \atom{CO}{}{}}$ and 19--25 for the  $\mathrm{GDR}$,
and the median values are $\alpha_{\rm \atom{CO}{}{}} = 0.5$~$M_{\Sol}$~pc$^{-2}$~(K~km~s$^{-1})^{-1}$ and  $\mathrm{GDR} = 20$, 
which will be used hereinafter.

\subsection{Measurements of the \C-to-\atom{H}{}{}$_2$ conversion factor $\alpha_{\rm \C}$}
\label{sec:alpha_c}
\begin{figure*}[tph]
 \begin{center}
  \includegraphics[width=0.8\linewidth]{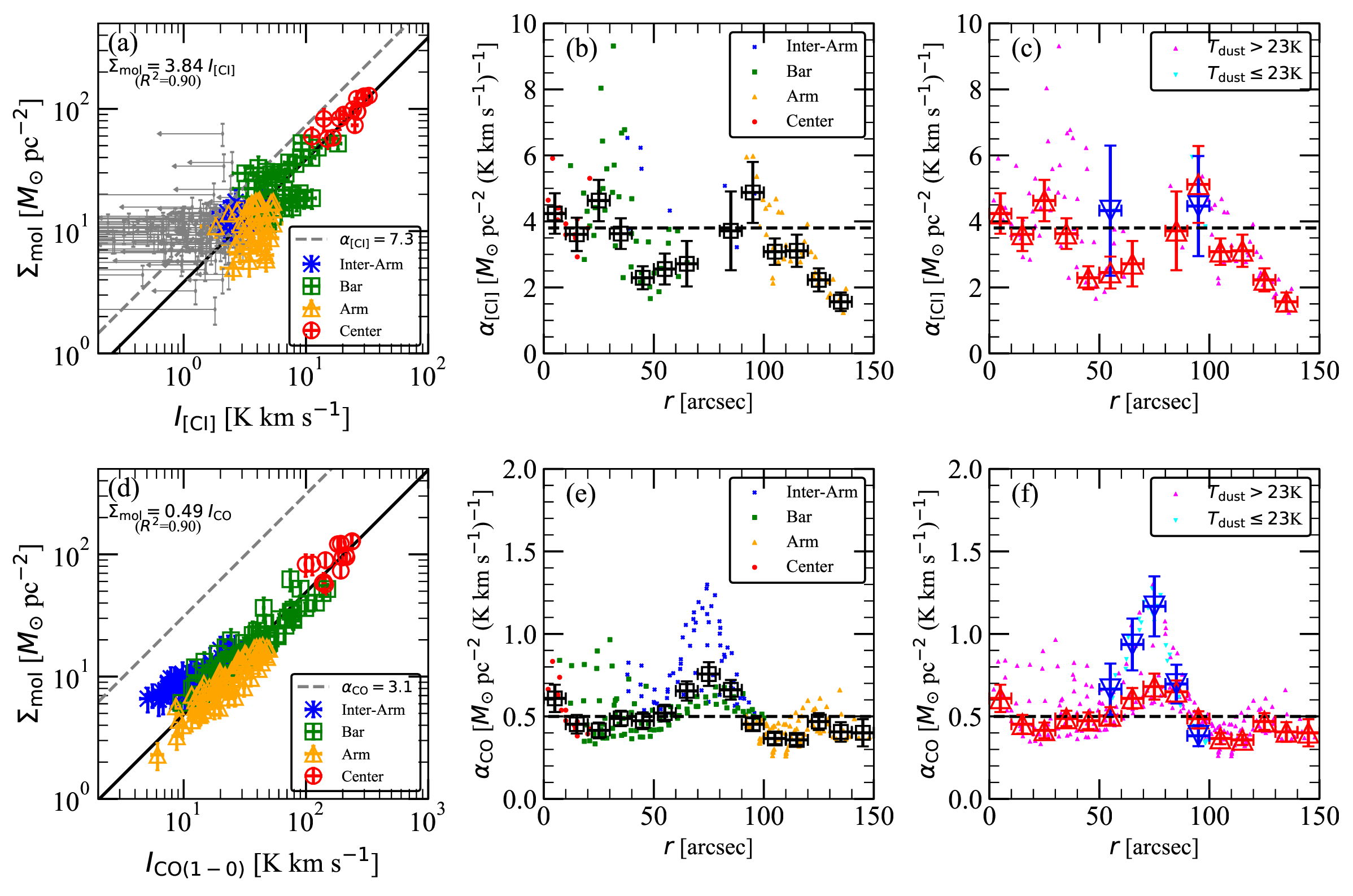}
 \end{center}
 \caption{  
 (a) Plots of $\Sigma_{\rm mol}$ as a function of $I_{\rm \C(1-0)}$, where the data with $I_{\rm \C(1-0)} < 4~\sigma$ are represented by gray points, and the remaining graphs are represented in the same manner as in Figure~\ref{fig:Dust-CICO}. 
  The best-fit relation with a fixed slope of unity to the data with $I_{\rm \C(1-0)} > 4~\sigma$ is shown by the  solid line, corresponding to $\alpha_{\rm \C}\sim3.8~[M_{\Sol}$~pc$^{-2}$~(K~km~s$^{-1}$)$^{-1}$]. 
  The dashed line is the mean value of the local galaxies (7.3~[$M_{\Sol}$~pc$^{-2}$~(K~km~s$^{-1}$)$^{-1}$]) measured by \citet{Crocker2019}. 
 In panels (b) and (c), the radial variation of $\alpha_{\rm \C}$  is plotted, but in (c), the data are separated  into two halves using a division of $T_{\rm dust}=23$~K.
The open boxes in  panel (b) represent the values binned with a width of $r = \timeform{10"}$. 
The upward and downward triangles in panel (c) represent the values binned with a width of $r = \timeform{10"}$ as well, but for $T_{\rm dust}>23$~K and $\leq 23$~K, respectively.
d) Plots of $\Sigma_{\rm mol}$ as a function of $I_{\rm \atom{CO}{}{}(1-0)}$. The best-fit relation with a fixed slope of unity is shown by the  solid line, corresponding to $\alpha_{\rm \atom{CO}{}{}}\sim0.5~[M_{\Sol}$~pc$^{-2}$~(K~km~s$^{-1}$)$^{-1}$].
The dashed line in panel (d) represents the averaged value $\alpha_{\rm \atom{CO}{}{}} \sim 3.1$~[$M_{\Sol}$~pc$^{-2}$~(K~km~s$^{-1}$)$^{-1}$] in nearby galaxies measured by \citet{Sandstrom2013}.
 Panels (e) and (f) are the same as panel (b), and (c), but reflect the data for \atom{CO}{}{}(1--0).
}
 \label{fig:alpha_ci}
\end{figure*}

Assuming that the $\mathrm{GDR}$ is constant across the observed region and that \ci~traces the molecular gas, we measured the conversion factor of \C~to the molecular gas surface density by using the following equation:
\begin{eqnarray}
\alpha_{\rm \C}I_{\rm \C} &=& \Sigma_{\rm mol}\nonumber\\
&=& \mathrm{GDR}~\Sigma_{\rm dust} - \Sigma_{\rm atom}.
\end{eqnarray}
Although the same method as that described in section~\ref{subsec:alpha_co} can be used to measure $\mathrm{GDR}$ and  $\alpha_{\rm \C}$, 
we note that the results obtained with this method are biased toward brighter \ci~(cf. Figure~\ref{fig:Dust}),  resulting in their overestimation.
Thus, we used $\mathrm{GDR}=20$, as obtained in section~\ref{subsec:alpha_co}.
Employing the $\mathrm{GDR}$ for the entire region, 
we derived  $\alpha_{\rm \C}$ by the least square fitting to the $\Sigma_{\rm mol}-I_{\rm \C}$ relation above the 4~$\sigma$ noise level for $I_{\rm \C(1-0)}$  using the above equation. 
The slope corresponding to $\alpha_{\rm \C}$ was $3.8\pm0.1$~$M_{\Sol}$~pc$^{-2}$~(K~km~s$^{-1})^{-1}$ with $R^2 = 0.90$ (black solid line in Figure~\ref{fig:alpha_ci}(a)).
The deviation from the fitting is however large at 
 $I_{\rm \C}\lesssim7$~K~km~s$^{-1}$, particularly in the arm region.
 Thus if the the central region data are used, the fitting gives  $\alpha_{\rm \C}=4.0\pm0.2$~$M_{\Sol}$~pc$^{-2}$~(K~km~s$^{-1})^{-1}$.
The conversion factors  $\alpha_{\rm \C}=7.3$~$M_{\Sol}$~pc$^{-2}$~(K~km~s$^{-1})^{-1}$ for nearby galaxies \citep{Crocker2019}  and $\alpha_{\rm \C}=4.9$~$M_{\Sol}$~pc$^{-2}$~(K~km~s$^{-1})^{-1}$ for 
(U)LIRGs \citep{Jiao2017} are  larger than our results, although within the scatter.
However,  it must be noted that the conversion factor $\alpha_{\rm \C}$ is generally determined using the molecular gas surface density 
based on $\alpha_{\rm \atom{CO}{}{}}$ (in addition, the intensity ratio of \atom{CO}{}{}(2-1) to \atom{CO}{}{}(1-0) in some cases) obtained from the literature.
Through a linear regression fit for 30 nearby galaxies on the \ci~integrated intensity and \atom{H}{}{}$_2$ column density that is derived from the radiative transfer models using multiple  \atom{CO}{}{} and \atom{CO}{}{13} lines, 
\citet{Israel2020} has reported a lower conversion factor, 
$X_{\C}=9\pm2\times10^{19}$~cm$^{-2}$~(K~km~s$^{-1}$)$^{-1}$, corresponding to $\alpha_{\rm \C}=1.9$~$M_{\Sol}$~pc$^{-2}$~(K~km~s$^{-1})^{-1}$,  
however, notably  
the average values for  individual galaxies is almost twice as high.
Moreover, \citet{Izumi2020} found  a comparable value of $\alpha_{\rm \C}=4.4$~$M_{\Sol}$~pc$^{-2}$~(K~km~s$^{-1})^{-1}$ in the central region ($r<70$~pc) of NGC~7469 without using $\alpha_{\rm \atom{CO}{}{}}$, where they compared the line flux luminosity of \ci~with $M_{\rm mol}$, which is obtained by subtracting the stellar mass and the black hole mass from the dynamical mass by assuming a negligible dark matter.
The consistency of $\alpha_{\rm\C}$ in a different spatial scale implies the usefulness of  \ci~as a molecular gas tracer if \ci~is bright, although further measurements 
are necessary.

The relationship between $I_{\rm \atom{CO}{}{}(1-0)}$ and $\Sigma_{\rm mol}$, which is derived using $\Sigma_{\rm dust}$ and $\mathrm{GDR} =20$ 
 is plotted in figure~\ref{fig:alpha_ci}(d).
The best fitted slope of $\alpha_{\rm \atom{CO}{}{}}=0.49(\pm0.01)$~$M_{\Sol}$~pc$^{-2}$~(K~km~s$^{-1})^{-1}$ is derived with 
 $R^2 = 0.9$, which is consistent with the value ($\alpha_{\rm \atom{CO}{}{}}=0.5$~$M_{\Sol}$~pc$^{-2}$~(K~km~s$^{-1})^{-1}$) obtained in section~\ref{subsec:alpha_co}.
Overall, $I_{\rm \atom{CO}{}{}(1-0)}$ is well correlated with $\Sigma_{\rm mol}$; however, 
the data from the disk region, especially from the inter-arm, deviate by more than $0.3$~dex from the value of $\alpha_{\rm \atom{CO}{}{}}=0.49$~$M_{\Sol}$~pc$^{-2}$~(K~km~s$^{-1})^{-1}$ relation, i.e., the value is larger than $\alpha_{\rm \atom{CO}{}{}}=0.49$~$M_{\Sol}$~pc$^{-2}$~(K~km~s$^{-1})^{-1}$.
It has been  reported that $\alpha_{\rm \atom{CO}{}{}}$ is a function of the galactic radius (e.g., \cite{Nakai1995}).
In Figure~\ref{fig:alpha_ci}(d), 
$\alpha_{\rm \atom{CO}{}{}}$ is plotted as a function of the galactic radius, with binning of a width of $r=\timeform{10"}$; 
however, no systematic radial dependence is observed.
To investigate the effect of the local excitation condition on $\alpha_{\rm \atom{CO}{}{}}$, we divided the data into two parts based on the dust temperature of 23~K 
(which is the dust temperature averaged over the galaxy)
in Figures~\ref{fig:alpha_ci}(c) and (f).
The blue bins represent lower dust temperatures ($T_{\rm dust}\lesssim23$~K), and the red bins represent higher temperatures ($T_{\rm dust}>23$~K).
It is clearly visible that the higher values of $\alpha_{\rm \atom{CO}{}{}}$ are distributed in the low dust temperature region.
Meanwhile, the optical depth of $\atom{CO}{}{}$,  $\tau_{\rm \atom{CO}{}{}}$,  can cause the variations in $\alpha_{\rm \atom{CO}{}{}}$, i.e., a small $\tau_{\rm \atom{CO}{}{}}$ leads to a large  $\alpha_{\rm \atom{CO}{}{}}$ (e.g., \cite{Nakai1995}). 
In fact, \citet{Crosthwaite2002}  reported that \atom{CO}{}{} in the inter-arm region of M83 was optically thin, based on the high intensity ratio of \atom{CO}{}{}(2--1)/\atom{CO}{}{}(1--0) $(>1)$.
Here, notably, $\tau_{\rm \atom{CO}{}{}}$ scales as $\tau_{\rm \atom{CO}{}{}}\propto \exp(h\nu/k T_{\rm ex})N_{\rm \atom{CO}{}{}}/\Delta v$, 
where $N_{\rm \atom{CO}{}{}}$ is the \atom{CO}{}{} column density and $\Delta v$ the velocity width (e.g., \cite{Bolatto2013}).
Considering the low temperature and narrow velocity width in the inter-arm region, a low \atom{CO}{}{} abundance may cause the low optical depth, and consequently the higher values of $\alpha_{\rm \atom{CO}{}{}}$.
The values of $\alpha_{\rm \C}$ do not show the systematic radial dependence nor a significant difference between low and high temperatures.
The dependences of $\alpha_{\rm \atom{CO}{}{}}$ and $\alpha_{\rm \C}$ on the dust temperature are consistent with the results shown by \citet{Crocker2019}.
However, 
only a  few \ci~data are available for low dust temperatures.

We compared the molecular gas mass ($M_{\rm mol, dust}$, $M_{\rm mol, CO}$, and $M_{\rm mol, \C}$) in the studied area by 
applying $\mathrm{GDR}=20$, $\alpha_{\rm \atom{CO}{}{}}=0.5$~$M_{\Sol}$~pc$^{-2}$~(K~km~s$^{-1})^{-1}$, and $\alpha_{\rm \C}=3.8$~$M_{\Sol}$~pc$^{-2}$~(K~km~s$^{-1})^{-1}$ to the dust surface density, including the atomic gas surface density and  the integrated intensities of \atom{CO}{}{}(1--0) and \ci. 
The resulting masses are summarized in Table~\ref{tab:mass}.
Although the mass in the entire area  is consistent within the error,  
\ci~tends to underestimate the mass due to non-detection in regions with low dust temperature such as the inter-arm region.

\begin{table*}
  \tbl{Molecular Gas Mass Estimated using Dust, \atom{CO}{}{}, and \C
  }{%
  \begin{tabular}{cccccc}
      \hline
      Mass[$10^7~M_{\Sol}$]					& 	 all 				& Center 			& Arm				& Bar 			& Interarm \\ 
      \hline
      $M_{\rm mol,  Dust} (\mathrm{GDR}=20)$ 	& $8.3\pm1.4$			&$2.2\pm0.3$		&$1.4\pm0.2$			&$2.7\pm0.5$		&$1.2\pm0.3$\\
      $M_{\rm mol, CO} (\alpha_{\rm \atom{CO}{}{}}=0.5)$	& $8.1\pm0.2$			&$2.1\pm0.1$		&$1.7\pm0.1$			&$2.8\pm0.1$		&$0.8\pm0.1$\\
      $M_{\rm mol, \C} (\alpha_{\rm \C} =3.8)$	& $7.0\pm1.9$			&$2.1\pm0.1$		&$1.5\pm0.4$			&$2.1\pm0.4$		&$0.6\pm0.3$\\
      \hline
    \end{tabular}}\label{tab:mass}
\end{table*}

\subsection{Excitation temperature, column density, and abundance of atomic carbon}
\label{subsec:NC}
\begin{figure}[th]
 \begin{center}
  \includegraphics[width=\linewidth]{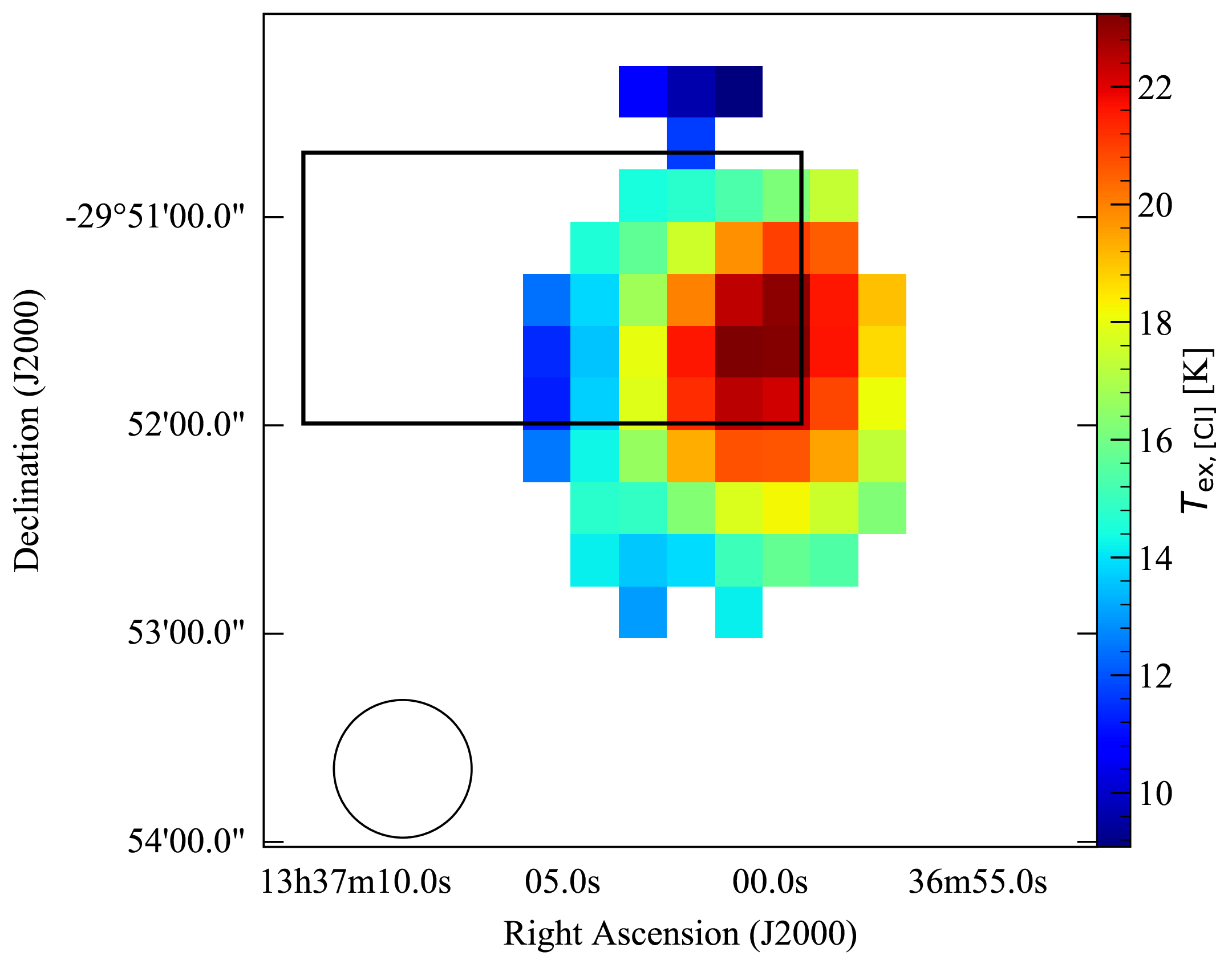}
 \end{center}
 \caption{ 
 Spatial distribution of $T_{\rm ex}$ derived by substituting the ratio between \C(1--0) and \C(2--1), retrieved  from the Very Nearby Galaxy Survey (VNGS) with Herschel/SPIRE FTS \citep{Wu2015}, into equation~(\ref{eq:R2}). 
 The region indicated by the box  represents the \ci~mapping area with ASTE.
}
 \label{fig:Tex}
\end{figure}
In this section, we derive the excitation temperature ($T_{\rm ex}$) and column density of  atomic carbon ($N_{\rm \atom{C}{}{}\emissiontype{I}}$), and the [\atom{C}{}{}\emissiontype{I}]/[\atom{H}{}{}$_2$] abundance ratio under the assumption that \atom{C}{}{}\emissiontype{I} is in local thermodynamic equilibrium (LTE) and the two \C~lines are optically thin.

The measured (background subtracted) radiation temperature from a region of uniform excitation temperature, $T_{\rm ex}$, is expressed as: 
\begin{eqnarray}
T_{\rm R}(T_{\rm ex}, \nu) =\eta \left[J(T_{\rm ex}, \nu) -J(T_{\rm CMB}, \nu) \right]\left(1-e^{-\tau}\right),
\label{eq:TR}
\end{eqnarray}
where $\eta$ is the beam filling factor, $J(T, \nu) = (h\nu/ k)/(e^{h\nu /kT}- 1)$ is the brightness temperature,  $T_{\rm CMB}$ is the cosmic microwave background temperature ($=2.73$~K), and $\tau$ is the optical depth. 
Assuming that the beam filling factor of \C(1-0) is similar to that of  \C(2-1) and $T_{\rm ex}=T_{\rm ex, \C(1-0)}=T_{\rm ex, \C(2-1)}$, 
the ratio between $T_{\rm R, \C(1-0)}$ and $T_{\rm R, \C(2-1)}$ is 
\begin{eqnarray}
R_{\rm \atom{C}{}{}\emissiontype{I}_{21}/\atom{C}{}{}\emissiontype{I}_{10}} &=& 
\frac{J(T_{\rm ex}, \nu_{\rm \atom{C}{}{}\emissiontype{I}_{21}}) - J(T_{\rm CMB}, \nu_{\rm \atom{C}{}{}\emissiontype{I}_{21}})}{J(T_{\rm ex}, \nu_{\rm \atom{C}{}{}\emissiontype{I}_{10}}) - J(T_{\rm CMB}, \nu_{\rm \atom{C}{}{}\emissiontype{I}_{10}})} \frac{(1-e^{-\tau_{\rm \atom{C}{}{}\emissiontype{I}_{21}}})}{(1-e^{-\tau_{\rm \atom{C}{}{}\emissiontype{I}_{10}}})} \nonumber\\
&=& \mathcal{K} \frac{(1-e^{-\tau_{\rm \atom{C}{}{}\emissiontype{I}_{21}}})}{(1-e^{-\tau_{\rm \atom{C}{}{}\emissiontype{I}_{10}}})}, 
\label{eq:K}
\end{eqnarray}
where $\mathcal{K}\equiv(J(T_{\rm ex}, \nu_{\rm \atom{C}{}{}\emissiontype{I}_{21}}) - J(T_{\rm CMB}, \nu_{\rm \atom{C}{}{}\emissiontype{I}_{21}}))/(J(T_{\rm ex}, \nu_{\rm \atom{C}{}{}\emissiontype{I}_{10}}) - J(T_{\rm CMB}, \nu_{\rm \atom{C}{}{}\emissiontype{I}_{10}}))$.
Under the assumption that two \C~ lines are optically thin ($\tau_{\rm \atom{C}{}{}\emissiontype{I}_{10}}, \tau_{\rm \atom{C}{}{}\emissiontype{I}_{21}}\ll1$), equation~(\ref{eq:K}) can be described as 
\begin{eqnarray}
R_{\rm \atom{C}{}{}\emissiontype{I}_{21}/\atom{C}{}{}\emissiontype{I}_{10}} &=& \mathcal{K}~\frac{\tau_{\rm \atom{C}{}{}\emissiontype{I}_{21}}}{\tau_{\rm \atom{C}{}{}\emissiontype{I}_{10}}}.
\label{eq:R}
\end{eqnarray}
The optical depth of the \C~line is given by 
\begin{eqnarray}
 \tau_{\rm ul} = \frac{\rm c^3}{8\pi\nu_{\rm ul}^3}
 N_{\rm \atom{C}{}{}\emissiontype{I}} \frac{g_{\rm u}}{Q} e^{-E_{\rm u}/kT_{\rm ex}} A_{\rm ul} \frac{1}{\Delta V} (e^{h \nu_{\rm ul}/kT_{\rm ex}} -1),
\label{eq:tau}
\end{eqnarray}
where 
the subscripts  $u$ and $l$ denote the upper and lower energy levels, 
$c$ is the speed of light, $\nu_{\rm ul}$ is the frequency corresponding to the transition, 
$N_{\rm  \atom{C}{}{}\emissiontype{I}}$ is the total column density of  \atom{C}{}{}\emissiontype{I}, 
$g_{\rm u}$ is the statistical weight,
$Q$ is the partition function,
$E_{\rm u}$ is the energy of the level (i.e., $E_1/k=23.6$~K and $E_2/k=62.5$~K for the $^3P_1$ and $^3P_2$ levels, respectively), 
$k$ is the Boltzmann constant, 
$A_{\rm ul}$ is the Einstein coefficient ($A_{10}=10^{-7.09725}$~s$^{-1}$ and $A_{21}=10^{-6.57415}$~s$^{-1}$, \cite{Muller2005}), 
and $\Delta V$ is the velocity width (see appendix~A of \citet{Salak2019} for more details of the derivation).
Substituting equation~(\ref{eq:tau}) into equation~(\ref{eq:R}) 
under the assumption that $\Delta V$ of \C(2--1) is the same as that of \C(1--0),
we obtain 
\begin{eqnarray}
R_{\rm \atom{C}{}{}\emissiontype{I}_{21}/\atom{C}{}{}\emissiontype{I}_{10}} &=& 
1.25~\mathcal{K}~\left(\frac{e^{-62.5/T_{\rm ex}}}{e^{-23.6/T_{\rm ex}}}\right)
\left(\frac{e^{h\nu_{21}/T_{\rm ex}}-1}{e^{h\nu_{10}/T_{\rm ex}}-1}\right).
\label{eq:R2}
\end{eqnarray}
Utilizing the integrated intensity ratio of \C(1--0) and \C(2--1), $T_{\rm ex}$  can be derived from equation~(\ref{eq:R2}).
We retrieved the \C(1--0) and \C(2--1) data of M83 from the VNGS with the Fourier transform spectrometer (FTS) of  Herschel/SPIRE \citep{Wu2015}, although the \C~data do not wholly cover our mapping area, e.g., the spiral arm.
Nevertheless, the archival data are convolved to a common spatial resolution of 41\farcs{7} and the pixels on the edge of each convolved map are truncated (cf. Figure~1 in \citet{Wu2015}).
The  $T_{\rm ex}$ distribution of M83 is shown in Figure~\ref{fig:Tex}, in which our mapping area is indicated by a black box.
The mean excitation temperature in our mapping area is $T_{\rm ex}=18\pm3$~K.
The temperature of $T_{\rm ex}\sim23$K in the central region  decreases with the radius and reaches to $\lesssim15$~K in the disk region.

The column density of atomic carbon gas $N_{\rm  \atom{C}{}{}\emissiontype{I}}$ in the LTE is given as
\begin{eqnarray}
N_{\rm  \atom{C}{}{}\emissiontype{I}} =\frac{8\pi\nu_{\rm 10}^3}{hc^3A_{10}} \frac{Q}{g_1}e^{E_1/T_{\rm ex}} I_{\rm \C}.
\label{eq:NC}
\end{eqnarray}
For $T_{\rm ex}=$13, 18, and 23~K,
the column densities in the mapping area are
$N_{\rm  \atom{C}{}{}\emissiontype{I}} = (7.4\pm1.5)\times10^{16}$~cm$^{-2}$, $(5.7\pm1.2)\times10^{16}$~cm$^{-2}$, and $(5.3\pm1.1)\times10^{16}$~cm$^{-2}$, respectively, that are obtained by using our integrated intensity of \ci.
$N_{\rm  \atom{C}{}{}\emissiontype{I}}$ for each region is summarized in Table~\ref{tab:abun}.
In addition, 
the [\atom{C}{}{}\emissiontype{I}]/[\atom{H}{}{}$_2$] abundance ratio can be derived 
using the \atom{H}{}{}$_2$ column density, 
$N_{\rm \atom{H}{}{}_2} (=4.59\times10^{19}~\Sigma_{\rm mol}$).
The mean abundance ratio in the mapping area is  $(6.9\pm1.8)\times10^{-5}$ for $T_{\rm ex}=18$~K.
This is comparable with the estimates of $\sim 7\times10^{-5}$ in the central region of the starburst galaxy NGC~1808 \citep{Salak2019} and $8.3\pm3.0\times10^{-5}$ for local (U)LIRGs \citep{Jiao2017}; however, our value is higher than the abundance of $2.5\pm1.0\times10^{-5}$ that is obtained as the average of the nearby galaxies that host starbursts and AGNs \citep{Jiao2019}.
Adopting  $T_{\rm ex}$ of 23, 18, 13~K for the central region, the disk (bar and arm) region, and the inter-arm region (Figure~\ref{fig:Tex}), respectively,  
the abundance ratio is found to be $\sim7\times10^{-5}$ within a range of 0.1~dex.

\begin{table*}
  \tbl{\atom{C}{}{}\emissiontype{I} Column densities and  Abundance Ratios of \atom{C}{}{}\emissiontype{I} and \atom{H}{}{}$_2$ in LTE
  }{%
  \begin{tabular}{ccccccc}
      \hline
      ~				& 	~	 			& all 				& Center 			& Bar				& Arm 			& Interarm \\ 
      \hline
      $T_{\rm ex} = 13$~K	& $N_{\rm C}$~[10$^{16}$~cm$^{-2}$]	&$7.4\pm1.5$		&$41.8\pm2.6$		& $7.2\pm1.5$			&$5.3\pm1.4$		&$2.7\pm1.5$\\
      ~  				& [C]/[H$_2$] ($\times10^{-5}$)			&$8.9\pm2.4$		&$9.9\pm1.5$		& $8.2\pm2.2$			&$11.4\pm3.6$		&$5.6\pm3.3$\\            
      \hline
      $T_{\rm ex} = 18$~K	& $N_{\rm C}$~[10$^{16}$~cm$^{-2}$]	&$5.7\pm1.2$		&$32.4\pm2.0$		& $5.6\pm1.1$			&$4.1\pm1.1$		&$2.1\pm1.1$\\
      ~  				& [C]/[H$_2$] ($\times10^{-5}$)			&$6.9\pm1.8$		&$7.7\pm1.2$		& $6.3\pm1.7$			&$8.9\pm2.8$		&$4.3\pm2.6$\\
      \hline
      $T_{\rm ex} = 23$~K	& $N_{\rm C}$~[10$^{16}$~cm$^{-2}$]	&$5.3\pm1.1$		&$29.8\pm1.8$		& $5.2\pm1.1$			&$3.8\pm1.0$		&$1.9\pm1.0$\\
      ~  				& [C]/[H$_2$] ($\times10^{-5}$)			&$6.3\pm1.7$		&$7.1\pm1.1$		& $5.8\pm1.6$			&$8.2\pm2.6$		&$4.0\pm2.4$\\      
      \hline
    \end{tabular}}\label{tab:abun}
\end{table*}

\subsection{Comparison of \ci~stacked intensity with dust temperature
}
\begin{figure}[th]
 \begin{center}
  \includegraphics[width=\linewidth]{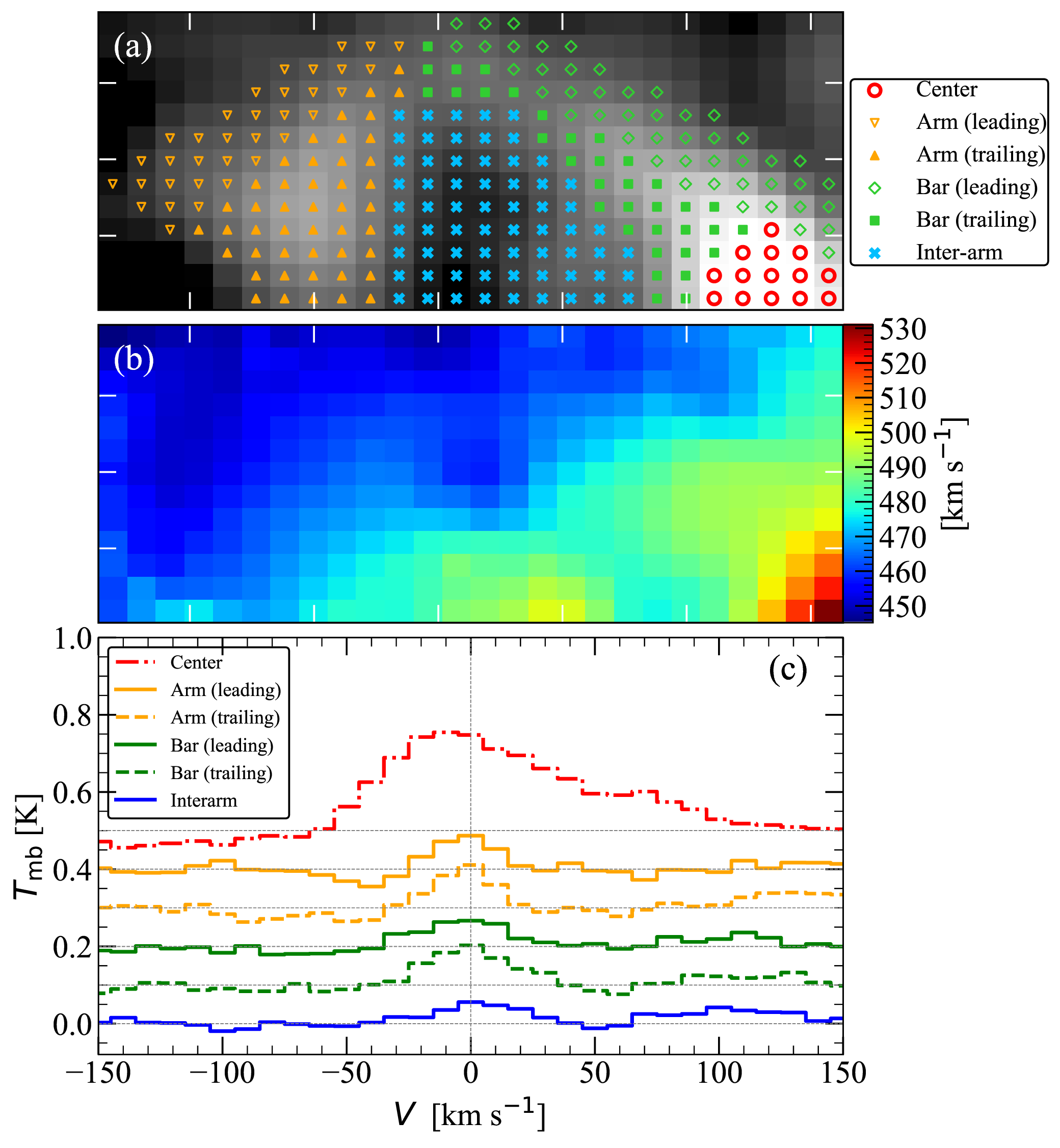}
 \end{center}
 \caption{ 
 (a) Integrated intensity maps of  \atom{CO}{}{}(1--0) overlaid with the region for the stacking analysis. 
 The center is represented by circles, leading side arm by downward triangles, trailing side arm by upward triangles, leading side bar by diamonds, trailing side bar by squares, and inter-arm by crosses.
(b) CO(1--0) velocity field map derived from intensity-weighted mean velocities.
(c) \ci~spectra  stacked in each galactic structure represented in (a). 
The vertical axis depicts the main-beam brightness temperature in K.
}
 \label{fig:stack_spe}
\end{figure}

\begin{figure}[th]
 \begin{center}
  \includegraphics[width=0.8\linewidth]{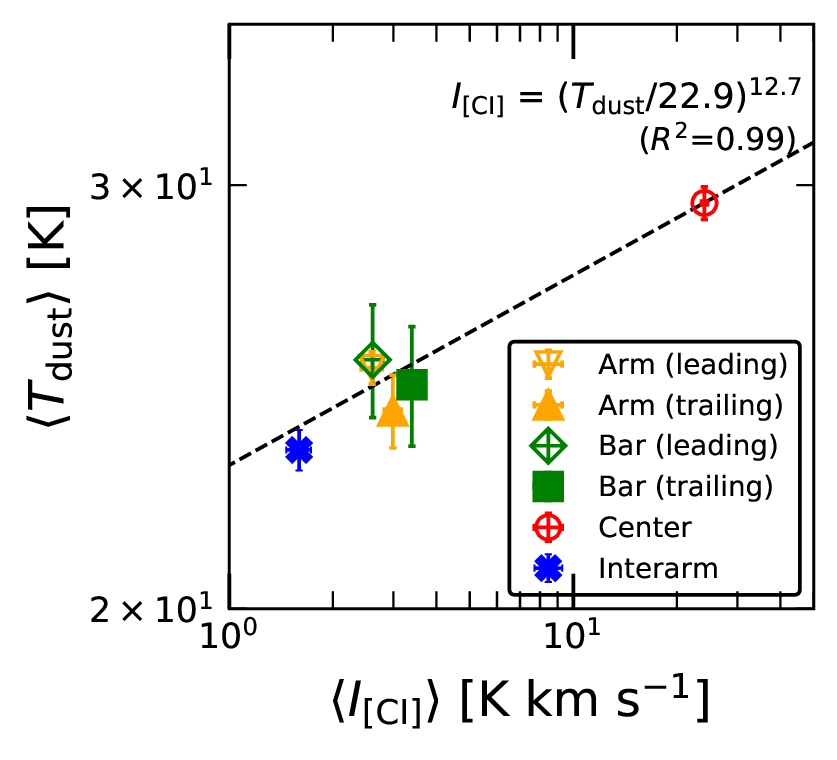}
 \end{center}
 \caption{
 Plot of $T_{\rm dust}$ averaged over each of the six regions represented in Figure~\ref{fig:stack_spe} as a function of the stacked integrated intensity of \ci, $\langle I_{\rm \C}\rangle$. 
 The dashed line  represents the best-fit relation of $\langle I_{\rm \C}\rangle = \left(\langle T_{\rm dust}\rangle/22.9\right)^{12.7}$.
 The symbols are the same as those in Figure~\ref{fig:stack_spe}.
 }
 \label{fig:cor_stack}
\end{figure}
In sections~\ref{sec:dust}, we have mentioned the possibility that \ci~is sensitive to local excitations; 
however,  much less \ci~data are available at a low $T_{\rm dust}$ region, such as in the inter-arm region.
To measure the dependence,
we employed a stacking technique (e.g., \cite{Morokuma2015}) using the intensity-weighted velocity field map of \atom{CO}{}{}(1--0) as a reference (Figure~\ref{fig:stack_spe}(b)).
We split the galaxy into six  characteristic  galactic structures: (1) center, (2) leading side of the arm, (3) trailing side of the arm, (4) leading side of the bar, (5) trailing side of the bar, and (6) inter-arm, as shown in Figure~\ref{fig:stack_spe}(a).
For the process, the \ci~spectra were first shifted along the velocity axis, according to the \atom{CO}{}{}(1-0) velocity field.
Next, the velocity-shifted spectra were averaged  in each region with equal weights for each velocity pixel. 
However,  this analysis is not very effective for \ci~if its velocity is different from that of \atom{CO}{}{}(1--0).

The resulting stacked \ci~spectra are shown in Figure~\ref{fig:stack_spe}.
\ci~line was detected in  all regions with S/N $>5$.
We examined the relation between the stacked spectral intensity ($\langle I_{\rm \C}\rangle$) and 
the dust temperature averaged over the corresponding regions ($\langle T_{\rm dust}\rangle$).
From figure~\ref{fig:cor_stack}, 
we find that the stacked intensity is a function of the dust temperature. 
The $\langle I_{\rm \C}\rangle$--$\langle T_{\rm dust}\rangle$ relationship is fitted by 
$\langle I_{\rm \C}\rangle = \left(\langle T_{\rm dust}\rangle / 22.9\pm0.4\right)^{12.7\pm0.9}$
with $R^2=0.99$.
The strong correlation between $\langle I_{\rm \C}\rangle$ and $\langle T_{\rm dust}\rangle$ indicates that \ci~is sensitive to the dust temperature, especially for $T_{\rm dust}\gtrsim23$~K (cf. $E/k=23.6$~K for the excitation energy of the upper level of \ci).
The dust temperature is associated with the external far ultraviolet (FUV) flux. 
Assuming that the FUV flux incident on a cloud surface is equal to the outgoing flux of dust radiation from the cloud, 
the UV radiation can be expressed as a function of $T_{\rm dust}$ in units of the Habing Field, i.e., $G_0= 1.6 \times 10^{-3}$~ergs~cm$^{-2}$~s$^{-1}$
(e.g., $G/G_0=(T_{\rm dust}/ 12. 2~{\rm K})^5$, \cite{Hollenbach1991}).
Classical PDR models have suggested that \atom{CO}{}{} is easily photodissociated by FUV photons, arising from massive stars (e.g., \cite{Tielens1985}, \cite{vandishoeck}, \cite{Hollenbach1991}, \cite{Kaufman1999}); 
hence, \atom{C}{}{}\emissiontype{I} predominantly exists in the surfaces of the UV-irradiated molecular clouds.
Meanwhile, the models have argued the weak dependence of the \ci~intensity on the dust temperature, which fails to explain our results.
We note that the limited range of $T_{\rm dust}$, measured by this work, may have resulted in the observed high exponential dependence of \ci~intensity.
To confirm this dependence, further measurements are required in a much wider range of $T_{\rm dust}$.

\bigskip
We demonstrated that \ci~was distributed outside the \atom{CO}{}{} gas in the spiral arm, although the \ci~distribution in the central region was similar to that of the \atom{CO}{}{}, which is consistent with the previous studies.
Moreover, the enhanced \ci~on the leading side (outer region) of the arm  was in good agreement with 
the warm dust traced by 70~\um,
atomic gas traced by \atom{H}{}{}\emissiontype{I},
OB stars traced by \atom{H}{}{}\emissiontype{$\alpha$},
and  interstellar radiation field indicated by the distribution of $T_{\rm dust}$.
These results suggest that the atomic carbon is a photodissociation product of \atom{CO}{}{}.
Consequently, \ci~is less reliable in tracing the bulk of "cold" molecular gas in the disk of the galaxy, although it is correlated with \atom{CO}{}{} in the central region.

Recent simulations have suggested that cosmic rays can induce the destruction of \atom{CO}{}{}, thus leaving behind \atom{C}{}{}\emissiontype{I}-rich molecular gas 
(e.g., \cite{Bisbas2015}, \cite{Glover2015}, \cite{Papadopoulos2018}).
Furthermore, the dust temperature was found to be inversely correlated with metallicity above $12+ \log(\atom{O}{}{}/\atom{H}{}{})\sim 8$, increasing from $T_{\rm dust}\sim22$~K near the solar metallicity ($12+ \log(\atom{O}{}{}/\atom{H}{}{})\sim 8.7$) to 35~K near a metallicity of 8 \citep{Engelbracht2008}.
If the reduced dust shielding in the cloud in the galaxies with low metallicity  causes \atom{CO}{}{} to be preferentially photo-dissociated relative to \atom{H}{}{}$_2$, the formation of \atom{C}{}{}\emissiontype{I} (and \atom{C}{}{}\emissiontype{II})-dominated H$_2$  is expected rather than \atom{CO}{}{}.
The increasing trend in the ratio of $I_{\rm \C}/I_{\rm \atom{CO}{}{}}$ for decreasing metallicity in the range of $12+ \log(\atom{O}{}{}/\atom{H}{}{})\sim8-9.1$ has been reported by \citet{Bolatto2000}.
Although \atom{CO}{}{} is a good tracer of the total cold molecular gas mass in galaxies, 
\ci~could be a substitute for \atom{CO}{}{} only in relatively hot regions, such as galactic centers and star-forming regions, 
and low-metallicity environments, as shown in this work and the previous studies 
(e.g., \cite{Israel2020}).

\section{Summary}
We have presented the image of the northern part of the spiral galaxy M83 in \ci~emission, 
and  compared the \ci~distribution with \atom{CO}{}{} and molecular gas traced by dust across the galactic structures, including the center, bar, arm, and inter-arm, for the first time.
The results of our study are summarized as follows: 

\begin{enumerate}
\item We observed the nearby galaxy M83 in \ci~with ASTE.
\ci~is detected  in the central as well as the bar and arm regions. 
The \ci~distribution in the central region is similar to the distributions of \atom{CO}{}{}(1--0), \atom{CO}{}{}(3--2), and \atom{CO}{}{13}(1--0).
In the arm region, \ci~is found to be in the leading side, which is consistent with the results of \hi~and 70~$\mu$m rather than \atom{CO}{}{}(1--0) and 250~$\mu$m.

\item The \ci~line luminosity is well correlated with the \atom{CO}{}{}(1--0), \atom{C}{}{13}O(1--0), and \atom{CO}{}{}(3--2) luminosities in the central region; however, it is weak or not correlated in the disk region. 
Although the relationship obtained using the data only in the central region is in good agreement with the values in the literature, 
the slope is steeper when all data including those of the central and disk regions are used.

\item We derived the distributions of  $T_{\rm dust}$ and  $\Sigma_{\rm dust}$ through the modified blackbody fitting
to the surface brightness at 70, 160, and $250~\mu$m.
The distribution of $T_{\rm dust}$  was nearly consistent with \C~in the arm and central regions but was offset from the leading edges of \atom{CO}{}{} and dust in the bar and arm regions. 
The distribution of $\Sigma_{\rm dust}$  was much more consistent with \atom{CO}{}{}(1--0) than with \ci.

\item 
We estimated the \atom{CO}{}{}-to-\atom{H}{}{}$_2$ conversion factor $\alpha_{\rm \atom{CO}{}{}}$ and gas-to-dust ratio $\mathrm{GDR}$ simultaneously to minimize the scatter in the  $\mathrm{GDR}$  values in the mapping region.
We calculated $\alpha_{\rm \atom{CO}{}{}}=0.5~M_{\Sol}$~pc$^{-2}$~(K~km~s$^{-1}$)$^{-1}$ and $\mathrm{GDR} =20$.
The value of $\alpha_{\rm \atom{CO}{}{}}$ is consistent with that derived from the radiative transfer model in the central region of M83 \citep{Israel2001}; however, it is   lower than the average $\alpha_{\rm \atom{CO}{}{}}=3.1~M_{\Sol}$~pc$^{-2}$~(K~km~s$^{-1}$)$^{-1}$ for 
the disks of nearby star-forming galaxies \citep{Sandstrom2013}.
In addition to the uncertainties of at least factor of 2 in $\mathrm{GDR}$ and $\alpha_{\rm \atom{CO}{}{}}$, 
the single-temperature modified black body fitting adopted by us could cause overestimation of the dust mass  in a high-temperature environment, yielding the low $\mathrm{GDR}$ and $\alpha_{\rm \atom{CO}{}{}}$.
Applying $\mathrm{GDR}=20$, the \C-to-\atom{H}{}{}$_2$ conversion factor $\alpha_{\C}$ = 3.8$~M_{\Sol}$~pc$^{-2}$~(K~km~s$^{-1}$)$^{-1}$ was derived through the least square fitting to the $\Sigma_{\rm mol}$--$I_{\rm \C}$ relation.
The molecular gas masses in the entire area  estimated  through dust, \atom{CO}{}{}, and \C~were consistent within error; however, \ci~tended to underestimate the mass in the low-$T_{\rm dust}$ regions, such as the inter-arm.

\item We derived the \C~excitation temperature, $T_{\rm ex}$, by using the two \C~lines ($^3P_1-^3P_0$ and $^3P_2-^3P_1$) obtained with the Herschel.
The temperature of $T_{\rm ex}\sim23$~K in the central region decreases with the radius and becomes less than $15$~K in the disk region.
Given the mean temperature $T_{\rm ex}\sim18$~K in our mapping area, the column density and abundance of \atom{C}{}{}\emissiontype{I}  were $N_{\rm \atom{C}{}{}}=(5.7\pm1.2)\times10^{16}$~cm$^{-2}$ and $[{\rm \atom{C}{}{}}]/[{\rm \atom{H}{}{}_2}]=(6.9\pm1.8)\times10^{-5}$, respectively.
Adopting an appropriate $T_{\rm ex}$ in each region (center, disk, arm, inter-arm) yielded a constant abundance of $\sim 7\times10^{-5}$ within a range of 0.1~dex in all regions.

\item By applying the stacking analysis with aligned velocity, 
by referring to the \atom{CO}{}{}(1--0) velocity field, the \ci~spectra in the central and disk regions (the leading and trailing sides of the arm and bar, and inter-arm) were detected.
The stacked intensity of \ci~is strongly correlated to $ T_{\rm dust}$.
This strong correlation between $\langle I_{\rm \C}\rangle$ and $\langle T_{\rm dust}\rangle$ indicates that \ci~is sensitive to the dust temperature.
In addition, a comparison of the distributions of \ci, \atom{CO}{}{}, $T_{\rm dust}$, and other lines (e.g., \hi~and \atom{H}{}{}\emissiontype{$\alpha$}) suggests that the atomic carbon is a photodissociation product of \atom{CO}{}{}.
Our results reveal that 
\ci~is less reliable in tracing the bulk of "cold" molecular gas in the disk of the galaxy, although
it could be a substitute for \atom{CO}{}{} only in regions with relatively warm and/or \ci-bright intensity, such as at the galactic center.

\end{enumerate}

\begin{ack}
We would like to thank the referee, Frank Israel, for valuable comments on the manuscript.
The ASTE telescope is operated by National Astronomical Observatory of Japan (NAOJ).
This paper makes use of the following ALMA data: ADS/JAO.ALMA\#2012.1.00762.S, ADS/JAO.ALMA\#2015.1.01593.S. ALMA is a partnership of ESO (representing its member states), NSF (USA), and NINS (Japan), together with NRC (Canada), MOST and ASIAA (Taiwan), and KASI (Republic of Korea), in cooperation with the Republic of Chile. The Joint ALMA Observatory is operated by ESO, AUI/NRAO and NAOJ.
The VNGS data was accessed through the Herschel Database in Marseille (HeDaM - http://hedam.lam.fr) operated by CeSAM and hosted by the Laboratoire d'Astrophysique de Marseille.
The National Radio Astronomy Observatory is a facility of the National Science Foundation operated under cooperative agreement by Associated Universities, Inc.
This research has made use of the NASA/IPAC Extragalactic Database (NED), which is operated by the Jet Propulsion Laboratory, California Institute of Technology, under contract with the National Aeronautics and Space Administration.
Data analysis was in part carried out on the Multi-wavelength Data Analysis System operated by the Astronomy Data Center (ADC), National Astronomical Observatory of Japan.
This work was supported by JSPS KAKENHI Grant Numbers JP20K04034, JP19H00702.
\end{ack}




\end{document}